\documentclass[aps,prl,twocolumn,reprint,floatfix, superscriptaddress]{revtex4-2}

\usepackage[normalem]{ulem}
\usepackage[utf8]{inputenc}
\usepackage{graphicx,amsmath,amssymb,braket, dsfont}
\graphicspath{{figures/}}
\usepackage{color}
\definecolor{darkgreen}{rgb}{0.0, 0.4, 0.26}

\usepackage{physics}

\definecolor{myblue}{rgb}{0.2,0.2,0.8}
\definecolor{myzard}{cmyk}{0,0,0.05,0}
\definecolor{mywhite}{rgb}{1,1,1}
\definecolor{mywhite}{rgb}{1,1,1}
\definecolor{myred}{rgb}{1,0.,0.3}
\usepackage[colorlinks=true,citecolor=myblue,linkcolor=myred]{hyperref}

\def\be{\begin{equation}}
\def\ee{\end{equation}}
\def\ba{\begin{align}}
\def\enda{\end{align}}
\def\bi{\begin{itemize}}
\def\ei{\end{itemize}}

\def\II{{\rm I}}

\def\dd{{\rm d}}

\def\PP{\mathbb{P}}

\def\rr{\mathbf{r}}
\def\RR{\mathbf{R}}
\def\kk{\mathbf{k}}
\def\qq{\mathbf{q}}

\def\RR{\mathbf{R}}
\def\R{\tilde{R}}
\def\rr{\mathbf{r}}
\def\pp{\mathbf{p}}
\def\PP{\mathbf{P}}
\def\PPP{\boldsymbol{\mathcal{P}}}
\def\kk{\mathbf{k}}
\def\qq{\mathbf{q}}
\def\q{\tilde{q}}
\def\GG{\mathbf{G}}
\def\ggg{\mathbf{g}}
\def\dd{\mathbf{d}}
\def\KK{\mathbf{K}}
\def\EE{\mathbf{E}}
\def\HH{\mathbf{H}}
\def\II{\mathcal{I}}

\def\kp{\mathbf{k}\cdot\mathbf{p}}

\begin{document}

\title{Photon-mediated interactions near a Dirac photonic crystal slab}
\author{E.P. Navarro-Barón}
\email{epnavarrob@unal.edu.co}
\affiliation{Grupo de Superconductividad y Nanotecnología, Departamento de Física, Universidad Nacional de Colombia, Ciudad Universitaria, K. 45 No. 26-85, Bogotá D.C., Colombia}
\affiliation{Grupo de Óptica e Información Cuántica, Departamento de Física, Universidad Nacional de Colombia, Ciudad Universitaria, K. 45 No. 26-85, Bogotá D.C., Colombia}
\author{H. Vinck-Posada}
\email{hvinckp@unal.edu.co}
\affiliation{Grupo de Superconductividad y Nanotecnología, Departamento de Física, Universidad Nacional de Colombia, Ciudad Universitaria, K. 45 No. 26-85, Bogotá D.C., Colombia}
\affiliation{Grupo de Óptica e Información Cuántica, Departamento de Física, Universidad Nacional de Colombia, Ciudad Universitaria, K. 45 No. 26-85, Bogotá D.C., Colombia}
\author{A. Gonz\'alez-Tudela}
\email{a.gonzalez.tudela@csic.es}
\affiliation{
 Institute of Fundamental Physics IFF-CSIC, Calle Serrano 113b, E-28006 Madrid, Spain
}

\begin{abstract}
Dirac energy-dispersions are responsible of the extraordinary transport properties of graphene. This motivated the quest for engineering such energy dispersions also in photonics, where they have been predicted to lead to many exciting phenomena. One paradigmatic example is the possibility of obtaining power-law, decoherence-free, photon-mediated interactions between quantum emitters when they interact with such photonic baths. This prediction, however, has been obtained either by using toy-model baths, which neglect polarization effects, or by restricting the emitter position to high-symmetry points of the unit cell in the case of realistic structures. Here, we develop a semi-analytical theory of dipole radiation near photonic Dirac points in realistic structures that allows us to compute the effective photon-mediated interactions along the whole unit cell. Using this theory, we are able to find the positions that maximize the emitter interactions and their range, finding a trade-off between them. Besides, using the polarization degree of freedom, we also find positions where the nature of the collective interactions change from being coherent to dissipative ones. Thus, our results significantly improve the knowledge of Dirac light-matter interfaces, and can serve as a guidance for future experimental designs.
\end{abstract}

\maketitle

\section{Introduction}

The Dirac energy spectrum of graphene is the source of many of its extraordinary electronic properties~\cite{castroneto09a}. This has triggered the quest to translate these energy dispersions to other systems, like photonic crystals (PhC)~\cite{joannopoulos97a}, by exploiting the analogy between the electronic and electromagnetic wave propagation~\cite{haldane08a,sepkhanov07a,zandbergen10a,Bahat-Treidel2010,zhang08a,bravo12a}. In this photonic context, these energy dispersions have already been predicted to lead to many non-trivial phenomena such as realizing topologically~\cite{haldane08a} or pseudo-diffusive~\cite{sepkhanov07a} photon transport, observing Klein-tunneling~\cite{Bahat-Treidel2010}, or enhancing the Purcell factor over large areas~\cite{bravo12a}. One of the latest additions to the exciting features of Dirac photonics has been the possibility of obtaining decoherence-free, long-range (power-law) interactions between emitters when many of them couple to these type of structures~\cite{Gonzalez-Tudela2018,Perczel2020a}.

Long-range interactions are instrumental for many quantum information and simulation applications. For example, they can be harnessed to induce long-distance entanglement~\cite{shahmoon13a}, to improve the speed of quantum state transfer protocols~\cite{eldredge17a,Kuwahara2020,Tran2020c,Tran2021b}, or to explore (non-) equilibrium phenomena in frustrated spin models~\cite{hauke13a,richerme14a,gong16b,maghrebi16a,koffel12a,vodola14a,kastner11a}. In free-space, photon-mediated interactions are naturally long-ranged (scaling with $1/r^{3(1)}$ in the near (far) field), but they are unavoidable accompanied by dissipation~\cite{lehmberg70a,lehmberg70b}. Photonic band-gaps can be used to cancel this dissipation~\cite{douglas15a,Gonzalez-Tudela2015b}, but at expense of exponentially attenuating the decay of the resulting interactions. Dirac energy dispersions in two~\cite{Gonzalez-Tudela2018,Perczel2020a} and three-dimensions~\cite{Gonzalez-Tudela2018a,Ying2019,Garcia-Elcano2020, Garcia-Elcano2021} have been recently pointed out as a way of avoiding this trade-off, showing how they lead to power-law decaying interactions without any associated dissipation thanks to the singular nature of the density of states around the Dirac points.

In the two-dimensional case~\cite{Gonzalez-Tudela2018,Perczel2020a}, these interactions have been so far predicted to have an overall scaling with a fixed $1/r$-decay with the distance, $r$, between emitters. However, these calculations were done either using toy-model coupled-resonator baths~\cite{Gonzalez-Tudela2018}, or fixing the emitter positions at high-symmetry points of the unit cell for the case of realistic structures~\cite{Perczel2020a}, thus limiting the generality of the results. In this work, we go beyond these analyses and derive a general theory for realistic structures that enables us to characterize the emergent photon-mediated interactions for emitters placed at any position of the unit cell. To illustrate its power, we apply this theory to the same Dirac photonic structure analyzed in Ref.~\citenum{Perczel2020a} and find several remarkable results. First, we study the positions that optimize the interaction strength beyond the one considered in Ref.~\citenum{Perczel2020a}. We characterize both the regions within the dielectric and the air holes, that can be used to optimize the coupling of solid-state~\cite{evans18a,lodahl15a} or natural~\cite{goban13a,thompson13a,Kim2019,Yu2019} atomic systems, respectively. Second, we find that certain positions display an effective longer-ranged decay exponent, $1/r^\gamma$, with $\gamma<1$, and study the trade-off between the strength of interactions and their range. This tunability is important because it opens the exploration of other long-range interacting models beyond the ones pointed by Refs.~\citenum{Gonzalez-Tudela2018,Perczel2020a}. Third, by playing with the polarization degree of freedom, we also find positions where the nature of the collective interactions change from being coherent to incoherent ones, which might lead to strong super/sub-radiant effects~\cite{dicke54a}. 

The manuscript is divided as follows: First, we introduce the formalism that connects the classical Green functions of the photonic structure with the effective quantum emitter interactions appearing in a master equation description of the emitters' dynamics. Second, we describe the Dirac-photonic structure that we will consider along this manuscript, and characterize its band-structure and the properties of their eigenmodes around the Dirac point. Then, we explain the basic ingredients of the theory we develop to analyze the effective emitter interactions, that are, the guided-mode expansion technique~\cite{andreani06a} and the $\kp$ method~\cite{Ming1994,Hui1994,Sipe2000}.  Next, we use our theory to study the position-dependence of the coupling strength, the range, and the coherent/incoherent nature of the photon-mediated interactions appearing in these systems. Finally, we will summarize our findings and point to other possible applications of our method.

\section{Photon-mediated interactions in photonic-crystals \label{sec:green}}

\begin{figure} [tb]
    \centering
    \includegraphics[scale=0.43]{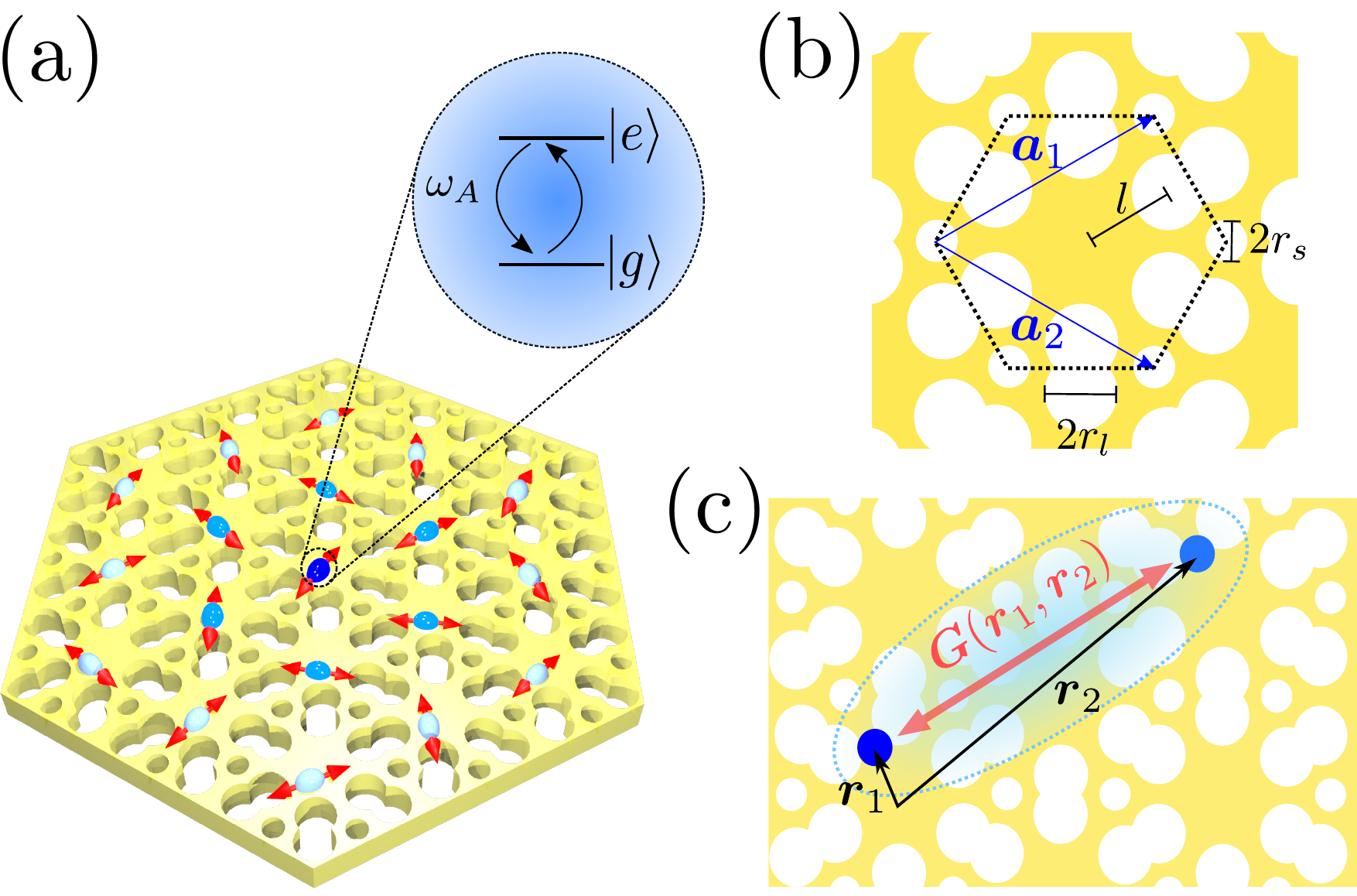}
    \caption{(a) Schematic 3D figure of the system: a collection of emitters are placed within or nearby a PhC slab with a thickness $d$. Each emitter has two levels, $\ket{g}$ and $\ket{e}$, with transition frequency $\omega_A$ between them. (b) Top view of the slab structure where we defined its geometric parameters: $r_l$ and $r_s$ are the radii of the inner and outer holes, $l$ is the distance between the center of the inner holes and the center of the unit cell, $\boldsymbol{a}_1$ and $\boldsymbol{a}_2$ are the primitive vectors for the hexagonal lattice. (c) The main focus of the work will be to obtain the two-point Green function, $\GG(\rr_1,\rr_2)$, which characterizes the interaction between emitters. We will restrict to situations where the emitters are placed at positions $\rr_1$ and $\rr_2=\rr_1+\RR$, being $\RR$ is a lattice displacement vector that can be written as a linear combination of the primitive vectors $\boldsymbol{a}_{1,2}$.}
    \label{fig: 1-system}
\end{figure}

The goal of this manuscript is to describe the emergent photon-mediated interactions when several quantum emitters couple to Dirac-like light-matter interfaces, as schematically depicted in Fig.~\ref{fig: 1-system}(a). A suitable formalism to describe these interactions in these dielectric media is macroscopic QED~\cite{Gruner1996, dung02a,buhmann07a,dzsotjan10a,Gonzalez-Tudela2011,Asenjo-Garcia2017a}, where the light-matter interactions are expressed in terms of the classical Green function, $\GG(\rr,\rr',\omega)$, obtained from the following electromagnetic wave equation:
\begin{equation}
    \nabla \times  \nabla \times \GG(\rr,\rr',\omega)+\frac{\omega^2}{c^2}\varepsilon(\rr,\omega)\GG(\rr,\rr',\omega)=\mathbb{1}\delta(\rr-\rr')\,,
\end{equation}
with $\varepsilon(\rr,\omega)$ being the permittivity of the medium. Within this formalism, and assuming that Born-Markov conditions are satisfied such that the photonic degrees of freedom can be adiabatically eliminated~\cite{Gruner1996,dung02a,buhmann07a,dzsotjan10a,Gonzalez-Tudela2011,Asenjo-Garcia2017a}, the resulting emitter dynamics can be described by the following master equation:

\begin{align}
\frac{d\rho(t)}{dt}=-\frac{i}{\hbar}[H_0+H_\mathrm{eff},\rho]+\mathcal{L}_\mathrm{eff}(\rho)\,, \label{eq:meq}
\end{align}
where, i) $H_0$ corresponds to the independent emitters' Hamiltonian. For this manuscript, we will assume that the emitters have a single optical transition from an optically excited state, $e$, to the ground state $g$, with frequency $\omega_A$ and dipole matrix element $\dd$, such that $H_0=\sum_{j}\hbar\omega_A\sigma^j_{ee}$, using the notation $\sigma^j_{\alpha\beta}=\ket{\alpha}_j\bra{\beta}$ for the emitter operators; ii) $H_\mathrm{eff}$ represents the unitary (coherent) part of the photon-mediated interactions that for two-level emitters reads:
\begin{align}
H_{\mathrm{eff}}=\sum_{i,j} J_{ij}\sigma_{eg}^i\sigma_{ge}^j\,.\label{eq:Hij}
\end{align}
Thus, this term yields coherent excitation exchanges between emitters at a rate $J_{ij}$, that can be harnessed, e.g., to make SWAP-like gates. Finally, iii) $\mathcal{L}_\mathrm{eff}(\rho)$ describes the non-unitary (incoherent) dynamics induced by the bath that reads:
\begin{align}
\mathcal{L}_\mathrm{eff}(\rho)=\sum_{ij}\frac{\Gamma_{ij}}{2}\left(2\sigma_{ge}^i\rho\sigma_{eg}^j-\sigma_{eg}^i\sigma_{ge}^j\rho-\rho\sigma_{eg}^i\sigma_{ge}^j\right)\,,\label{eq:Gammaij}
\end{align}
and accounts for both the individual ($\Gamma_{ii}$) and collective ($\Gamma_{i\neq j}$) dissipation. Despite being non-unitary, these collective dissipative terms yield strong super/sub-radiance effects~\cite{dicke54a}, which can be harnessed to engineer decoherence-free quantum gates~\cite{Paulisch2016,kockum18a} or to improve multi-photon generation~\cite{Gonzalez-Tudela2015,Gonzalez-Tudela2017,Paulisch2019} and absorption~\cite{Asenjo-Garcia2017a} fidelities, among other applications. Remarkably, both the coherent and incoherent terms of the master equation are related to the classical Green function as follows~\cite{dung02a,buhmann07a,dzsotjan10a,Gonzalez-Tudela2011,Asenjo-Garcia2017a}:
\begin{align}
J_{ij}&=-\frac{\mu_0\omega_A^2}{\hbar}\dd^{*}_i\cdot\mathrm{Re}[\GG(\rr_i,\rr_j,\omega_A)]\cdot\dd_j\,, \\
\Gamma_{ij}&=\frac{2\mu_0\omega_A^2}{\hbar}\dd^{*}_i\cdot \mathrm{Im}[\GG(\rr_i,\rr_j,\omega_A)]\cdot\dd_j\,,
\end{align}
where $\dd_i,\rr_i$ are the optical dipole moment and position of the $i$-th emitter, respectively. Thus, to know the photon-mediated interactions, $J_{ij},\Gamma_{ij}$ induced by a photonic media, it suffices to calculate the two-point Green function, $\GG(\rr_i,\rr_j,\omega_A)\equiv \GG(\rr_i,\rr_j) $, of the particular structure. Before explaining the method we develop to do it, let us first introduce the particular photonic structure we will use to benchmark our results. 

\section{Dirac photonic structure\label{sec:system}}

The photonic structure that we will consider is the photonic-crystal (PhC) slab depicted in Fig.~\ref{fig: 1-system}, which was first introduced in Ref.~\citenum{Perczel2020a}. The slab consists of a hexagonal lattice of GaP ($\varepsilon_{\mathrm{GaP}}=10.5625$) with primitive vectors $\mathbf{a}_{1/2}=(\sqrt{3}/2,\pm1/2)a$, being $a$ the lattice constant, and with six inner and six outer air holes, of radii $r_l$ and $r_s$, respectively. The unit cell of the PhC slab is shown in detail in Fig.~\ref{fig: 1-system}(b), where $l$ denotes the distance between the center of the unit cell and the inner circles, and $d$ the slab thickness. As shown in Ref.~\citenum{Perczel2020a}, an interesting property of this structure is that it features energetically isolated Dirac-cone dispersions at the high symmetry points $\KK=\frac{2\pi}{3a}(\sqrt{3},1)$ and $\KK'=\frac{2\pi}{3a}(\sqrt{3},-1)$ for certain parameters regime that are the ones we will focus along this manuscript.

\begin{figure}[tb]
    \centering
    \includegraphics[scale=0.37]{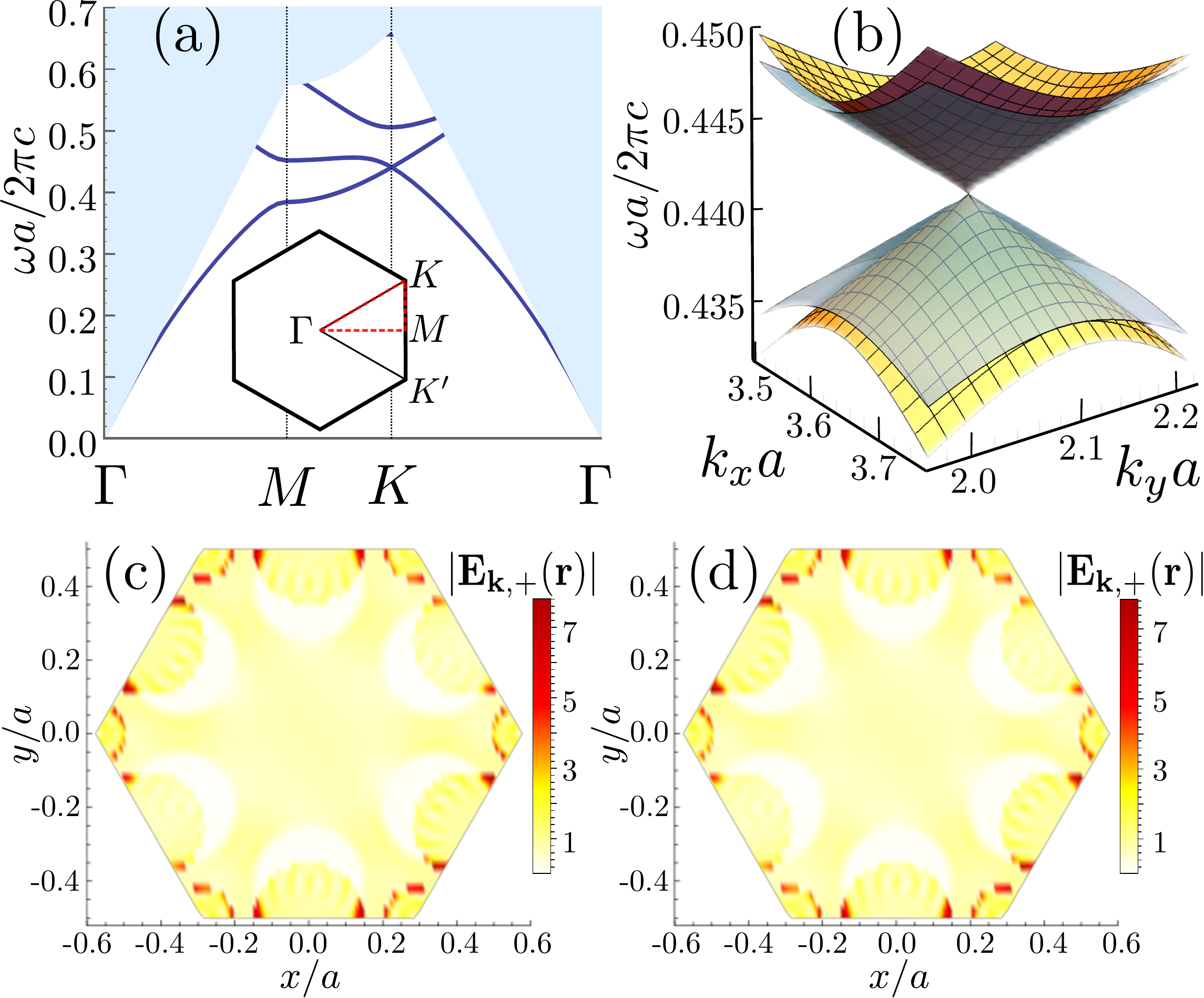}
    \caption{Band structure of the PhC slab of Fig.~\ref{fig: 1-system}. In panel (a), we show the numerical calculation of the GME method for dispersion around for a particular path of the first Brillouin zone, depicted in the inset. In panel (b), we compare the $\kp$ approximation with the GME method around the Dirac cone at $\KK$, where the light-blue surface correspond to the $\kp$-approximation and the yellow-orange tones correspond to the GME calculations. (c-d) Electric fields in the whole unit cell for an eigenmode in the upper band with momentum $\kk=\KK+\qq$, with  $|\qq|=0.03|\KK|$ and $\phi_\qq=0$. In the two panels we compare the results obtained numerically with GME method (c) and the fields obtained by the $\kp$ approximation (d).  The geometric parameters for these calculations are: $r_s=0.0833 a$, $r_l=0.15 a$,  $l=0.4 a$, $d=0.25a$; we use $1165$ different $\mathbf{g}$ vectors and consider $3$ guided modes for each $\mathbf{g}$, so we have a basis with total $3480$ guided modes.}
    \label{fig: 2-bands-cone}
\end{figure}

In Ref.~\citenum{Perczel2020a} the numerical analysis of the structure was done using a plane-wave expansion method~\cite{johnson01a}. With the results from these numerical calculations, they proposed an empirical ansatz of the band-structure ($\omega^{(n)}_\kk$) and their associated eigenmodes at the central position of the unit cell ($\EE_{\kk,n}(\rr=0)$) around the Dirac point $\KK$.  Then, using that ansatz they were able to construct the two-point Green function, $\GG(\rr_i,\rr_j)$ for emitters placed at the central point of the unit cell.
Here, we will use an alternative approach that consists on a combination of the Guided-Mode Expansion (GME) technique~\cite{andreani06a} and the $\kp$ method~\cite{Ming1994,Hui1994,Sipe2000}. The former allows one to find efficient solutions for the eigenfrequencies and eigenmodes of slab geometries, whereas the later will enable us to systematically expand the energy dispersions and eigenmodes around the Dirac point, and find the associated electromagnetic field \emph{at all positions} of the unit cell. This will be the key advantage of our approach, since we will be able to compute the two-point Green function, and thus the emergent photon-mediated interactions, for emitters placed at any point of the unit cell.

Let us now briefly explain the main steps of our theory, and apply it to characterize the photonic structure of Fig.~\ref{fig: 1-system}. The first step consists in using the GME technique~\cite{andreani06a}. This method initially finds the solutions of the homogeneous slab waveguide with an effective permittivity, that we denote by $\boldsymbol{h}_{\ggg,\mu}(\rr,z)$. Using Bloch-theorem we can rewrite these solutions as $\boldsymbol{h}_{\ggg,\mu}(\rr,z)=e^{i\kk\cdot\rr}\mathbf{U}_{\ggg,\mu}(\rr,z)$, separating the part $e^{i\kk\cdot\rr}$ given by Bloch theorem, from $\mathbf{U}_{\ggg,\mu}(\rr,z)$, that provides the real space distribution within each unit cell associated to each guided mode with momentum $\ggg$, and which has the same periodicity than the original PhC lattice. Here, $\ggg=\kk+\GG$ is a combination of $\kk$ inside the first Brillouin zone, $\GG$ a vector of the reciprocal lattice with information about the periodicity, and $\mu$ denotes the different modes of the effective homogeneous waveguide for a given $\ggg$, which is related with the $z$-quantization of the modes. Note also that here we use $\rr$ to express a vector in the $xy$ plane and we will write the dependence with the $z$ coordinate when needed. Using that form for the eigenmodes of the homogeneous slab, the magnetic field of the complete photonic structure is expanded in terms of these guided modes of the homogeneous waveguide (see Supporting Information), as  follows
\begin{equation}
    \mathbf{H}_{\kk,n}(\rr,z)=\sum_{\GG,\mu} c_n(\kk+\GG,\mu) \boldsymbol{h}_{\kk+\GG,\mu}(\rr,z)\, ,
    \label{Ec: 1-GME MagneticField}
\end{equation}
where $c_n(\kk+\GG,\mu)$ are the coefficients of expansion, and $n$ denotes the different modes of the PhC slab that appear with frequency $\omega^{(n)}_{\kk}$. Inputting this expansion into the eigenvalue problem of the Maxwell equations, one arrives to a linear matrix eigenvalue problem that can be solved numerically by truncating the size of the guided mode basis. Since we are interested in studying emitters placed at the $z=0$ plane of the slab, we can restrict to the study of transverse-electric TE-like (even) modes, because transverse-magnetic TM-like ones have an electric field zero at $z=0$ and thus will not couple to the emitter. 

In Fig.~\ref{fig: 2-bands-cone}(a) we plot the band-structure of the TE-like modes of the structure of Fig.~\ref{fig: 1-system} using the following geometric parameters: $r_s=0.0833 a$, $r_l=0.15 a$,  $l=0.4 a$, $d=0.25a$, and calculated with the GME method using a truncation for the homogeneous waveguide basis of $3480$ guided modes. There, we observe how indeed energetically isolated Dirac-cone dispersions at the $\KK$-point appear, that will be the main focus of this article. The next step consists in making a perturbative expansion around the $\KK$-point, i.e., $\kk=\KK+\qq$, for small $\qq$, using the $\kp$ method. The key idea of the perturbative expansion is to use the semi-analytical solutions obtained from the GME method at the $\KK$ point as the basis to express the problem at $\kk=\KK+\qq$. For that, we use agin the fact that the magnetic field within the PhC also satisfies Bloch theorem such that we can rewrite it also as  $\mathbf{H}_{\kk,n}(\rr,z)=e^{i\kk\cdot\rr}\boldsymbol{u}_{\kk,n}(\rr,z)$, where $\boldsymbol{u}_{\kk,n}(\rr,z)$ is the Bloch periodic function which contains the dependence of the electric field within the unit cell. Using the GME results, we can obtain the basis of periodic function $\{\boldsymbol{u}_{\KK,n}(\rr,z)\}$ around $\KK$ as follows:
\begin{equation}
    \boldsymbol{u}_{\KK,n}(\rr,z)=\sum_{\GG,\mu}c_n(\KK+\GG,\mu)\mathbf{U}_{\KK+\GG,\mu}(\rr,z)\, ,
    \label{ec:PeriodicFunction}
\end{equation}
with which we can write the magnetic field for momenta around the Dirac point, $\kk=\KK+\qq$, as 
\begin{equation}
    \HH_{\kk,n}(\rr,z)=e^{i (\KK+\qq)\cdot\rr}\sum_{j}C_{j}(\kk,n)\boldsymbol{u}_{\KK,j}(\rr,z)\, , \label{Ec:k.P MagneticField}
\end{equation}
Putting this expansion into the eigenproblem of the magnetic field, one arrives in the following equation
\begin{equation}
    \sum_{j}\mathsf{H}_{l,j} C_{j}(\kk,n)=\frac{\omega_{\kk,n}^{2}}{c^{2}}C_{l}(\kk,n)\, ,
    \label{ec:k.P general eigenproblem}
\end{equation}
where the matrix elements $\mathsf{H}_{l,j}$ can have terms of order $\qq^0$, $\qq^1$ to $\qq^2$, as explicitly shown in the Supporting Information. Since the Dirac points $\KK^{(')}$ have degenerate modes, we only consider the contribution of the two first bands $(l,j=1,2)$. Additionally, as we find that the energy dispersion is approximately linear, we can also neglect the order $\qq^2$ terms. With these considerations the general eigenvalue problem of Eq.~\eqref{ec:k.P general eigenproblem} reduces to a $2\times2$ matrix eigenproblem:
\begin{equation}
    \begin{pmatrix}
        \qq\cdot\PP_{11}&& \qq\cdot\PP_{12}\\
        \qq\cdot\PP^{*}_{12}&& -\qq\cdot\PP_{11}
    \end{pmatrix}\begin{pmatrix}
        \xi_{\pm}\\
        \eta_{\pm}
    \end{pmatrix}=\Delta\lambda_{    \pm}\begin{pmatrix}
        \xi_{\pm}\\
        \eta_{\pm}
    \end{pmatrix}
    \label{ec:k.p-MatrixEigenproblem}
\end{equation}
where $\Delta\lambda_{\pm}=\frac{\omega_{\kk\pm}^{2}-\omega_{D}^2}{c^{2}}$. The subindex $\pm$ is related to the upper and lower bands of the cone, and $\omega_{D}$ corresponds to the frequency at the Dirac point, which we define numerically as the average between $\omega_{\KK^{(')}}^{(1)}$ and $\omega_{\KK^{(')}}^{(2)}$. These particular relations between the matrix elements result from the crystal symmetry called ``deterministic degeneration"~\cite{Mei2012}. Details about $\PP_{l,j}$ and its calculation are given in the Supporting Information.

Solving the simplified eigenvalue problem of Eq.~\eqref{ec:k.p-MatrixEigenproblem}, we indeed obtain a linear dependence with $|\qq|$ of the eigenvalues, as $\Delta\lambda_{\pm}=\pm 2 \frac{v}{c}\frac{\omega_D}{c}|\qq|$, with $v$ being the group velocity at the Dirac points, with which we can approximate the energy dispersion of the two bands as follows:
\begin{align}
\omega_\pm(\KK^{(')}+\qq)\approx \omega_D\pm v |\qq|  \,.\label{eq:wkaprox}
\end{align}

In Fig.~\ref{fig: 2-bands-cone}(b) we plot these analytical approximations (in shaded blue surface) together with the numerically obtained energy bands using the GME method (orange surface), showing a good agreement between the two. Interestingly, from the simplified eigenvalue problem obtained through the $\kp$ approximation, we can also obtain the following magnetic and electric field expansions around the Dirac points
\begin{eqnarray}
\mathbf{H}_{\kk,\pm}(\rr,z)&\approx e^{i \qq\cdot\rr}\left[\xi_{\pm}\mathbf{H}_{\kk_{0},1}(\rr,z)+\eta_{\pm}\mathbf{H}_{\kk_{0},2}(\rr,z)\right] \, ,\\
\mathbf{E}_{\kk,\pm}(\rr,z)&\approx e^{i \qq\cdot\rr}\left[\xi_{\pm}\mathbf{E}_{\kk_{0},1}(\rr,z)+\eta_{\pm}\mathbf{E}_{\kk_{0},2}(\rr,z)\right]\label{Ec:  6-Efield_with_kP1}\, ,
\end{eqnarray}
where $\kk_0$ can be $\KK$ or $\KK'$; and the subindices $1,2$ indicate the two degenerate modes at $\KK^{(')}$. The parameters $\xi_\pm$ and $\eta_\pm$ have the following values for $\KK^{(')}$:
\begin{align}
    \xi_+=&\sin\left[\frac{\phi_\qq-\delta_{\KK^{(')}}}{2}\right]\, ,\\
    \eta_+=&\mp\cos\left[\frac{\phi_\qq-\delta_{\KK^{(')}}}{2}\right]\, ,\\
    \xi_-=&\pm\cos\left[\frac{\phi_\qq-\delta_{\KK^{(')}}}{2}\right]\, ,\\
    \eta_-=&\sin\left[\frac{\phi_\qq-\delta_{\KK^{(')}}}{2}\right]\, ,
\end{align}
where the symbols $\pm$ and $\mp$ are for $\KK/\KK'$, respectively, $\tan(\phi_\qq)=q_y/q_x$, and the phases $\delta_{\KK^{(')}}=\pm\pi/6$. In Figs.~\ref{fig: 2-bands-cone}(c-d) we plot a comparison of the electric field amplitude corresponding to a particular $\kk$ value of the upper band along the whole unit cell obtained numerically through the GME method (c) and semi-analytically using the $\kp$-approximation (d). There, we can see how the approximation captures indeed very well the emergent physics along the whole unit cell. Finally, let us also note that despite the different appearance of the expressions of the electric field in Eq.~\eqref{Ec:  6-Efield_with_kP1} with respect to the empirical ansatz proposed in Ref.~\citenum{Perczel2020a} for $\rr=0$, they reproduce the same physics if we restrict our approximated fields to this position, as we explicitly show in Supporting Information.

\section{Constructing Green functions}

Now, we will use the analytical approximations of Eqs.~\eqref{eq:wkaprox}-\eqref{Ec:  6-Efield_with_kP1} to construct the Green function of the problem, as also did in Ref.~\citenum{Perczel2020a}. However, the advantage of our method is that the electric field expansion of Eq.~\eqref{Ec: 6-Efield_with_kP1} is valid for any position of the unit cell, not only for the center $\rr=0$, and thus, we will be able to calculate $\GG(\rr_1,\rr_2)$ in a more general fashion. For this work, we restrict to the situation in which the position of the two emitters differs only by primitive lattice displacement, that is, $\rr_2=\rr_1+\RR$, where $\RR=\sum_{i=1}^2 a_i\mathbf{a}_i$. Thus, we only target to calculate: $G_{\alpha\beta}(\rr_1,\rr_1+\RR):=G_{\alpha\beta}(\rr_1;\RR)$. This Green function can be calculated integrating the momentum-space Green function as follows:
\begin{align}
G_{\alpha\beta}(\rr_1;\RR)=\iint_\mathrm{BZ} \frac{d^2\pp}{(2\pi)^2} g_{\alpha\beta}(\pp)\,,\label{eq:intGij}
\end{align}
with:
\begin{align}
g_{\alpha\beta}(\pp)=\frac{\sqrt{3}a^2}{2}c^2\sum_{n}\frac{\EE^{(n)*}_{\pp,\alpha}(\rr_1)\EE^{(n)}_{\pp,\beta}(\rr_1+\RR)}{\omega_A^2-\left(\omega^{(n)}_{\pp}\right)^2}
\end{align}
where $\EE^{(n)}_{\pp,\alpha}(\rr_i)$ is the electric-field $\alpha$-component at the $\rr_i$ position associated to the eigen-energy $\omega_\pp^{(n)}$ of the $n$-th band. Note that due to the periodicity of the PhC (see Eq.~\eqref{Ec:  6-Efield_with_kP1}), $\EE^{(n)*}_{\pp,\alpha}(\rr_1)\EE^{(n)}_{\pp,\beta}(\rr_1+\RR)\propto u^{(n)*}_{\pp,\alpha}(\rr_1) u^{(n)}_{\pp,\beta}(\rr_1)e^{-i\pp \cdot\RR}$. Assuming that $\omega_A$ is close to $ \omega_D$, then $\omega_A^2-\left(\omega^{(n)}_{\pp}\right)^2\approx2\omega_A (\omega_A-\omega^{(n)}_{\pp})$, and the integral in Eq.~\eqref{eq:intGij} is mostly given by the contributions around the $\KK^{(')}$-points. Then, we can use the analytical approximations of Eqs.~\eqref{eq:wkaprox}-\eqref{Ec:  6-Efield_with_kP1} to obtain the Green-function components:
\begin{align}
G_{\alpha\beta}(\rr_1;\RR)\approx& \frac{\sqrt{3}a^2c^2\delta_A}{16\, \omega_A v^2}\left\lbrace i H_0^{(1)}(R\delta_A/v)\times\right. \nonumber \\
& \left[\mathcal{A}^{\KK}_{\alpha\beta}(\rr_1) e^{i\KK\cdot\RR}+\mathcal{A}^{\KK'}_{\alpha\beta}(\rr_1) e^{i\KK'\cdot\RR}\right] \nonumber\\
&-H_1^{(1)}(R\delta_A/v)\mathcal{B}^{\KK}_{\alpha\beta}(\rr_1,\phi)e^{i\KK\cdot\RR}\nonumber\\
&\left.-H_1^{(1)}(R\delta_A/v)\mathcal{B}^{\KK'}_{\alpha\beta}(\rr_1,\phi)e^{i\KK'\cdot\RR}·\right\rbrace
~\label{eq:Gab}
\end{align}
where $R=|\RR|$, $\tan(\phi)=R_y/R_x$, $\delta_A=\omega_D-\omega_A$, $H_j^{(1)}(r)$ are the first kind Hankel function of order $j$, and $\mathcal{A}^{\KK^{(')}}_{\alpha\beta}(\rr_1),\,\mathcal{B}^{\KK^{(')}}_{\alpha\beta}(\rr_1,\phi)$ are coefficients defined in terms of the direction of $\RR$ and the electrical fields at position $\rr_1$ for the two first modes of $\KK^{(')}$. More details about the explicit form of these coefficients can be found in the Supporting Information. We do not consider $G_{\alpha z}$ because it is strictly zero for emitters placed at the symmetry plane $z=0$. Finally, let us note that the Green functions associated to the circular polarized components can be reconstructed from Eq.~\eqref{eq:Gab} as follows:
\begin{equation}
G_{\sigma_{\pm}\sigma_{\pm}}(\rr_1;\RR)=\frac{G_{yy}\mp iG_{xy}\pm iG_{yx}+G_{xx}}{2}\label{eq:Gsp(m)sp(m)}
\end{equation}
\begin{equation}
    G_{\sigma_{\pm}\sigma_{\mp}}(\rr_1;\RR)=\frac{G_{yy}\mp iG_{xy}\mp iG_{yx}-G_{xx}}{2}\label{eq:Gsp(m)sm(p)}
\end{equation}
In the next section, we analyze these functions in detail, putting special emphasis on the dependence of $G_{\alpha\beta}(\rr_1,\RR)$ at different places of the unit cell, since it is the main strength of our method.

\section{Dirac-Photon-mediated interactions along the whole unit cell \label{sec:results}}

\begin{figure}[tb]
    \centering
    \includegraphics[scale=0.5]{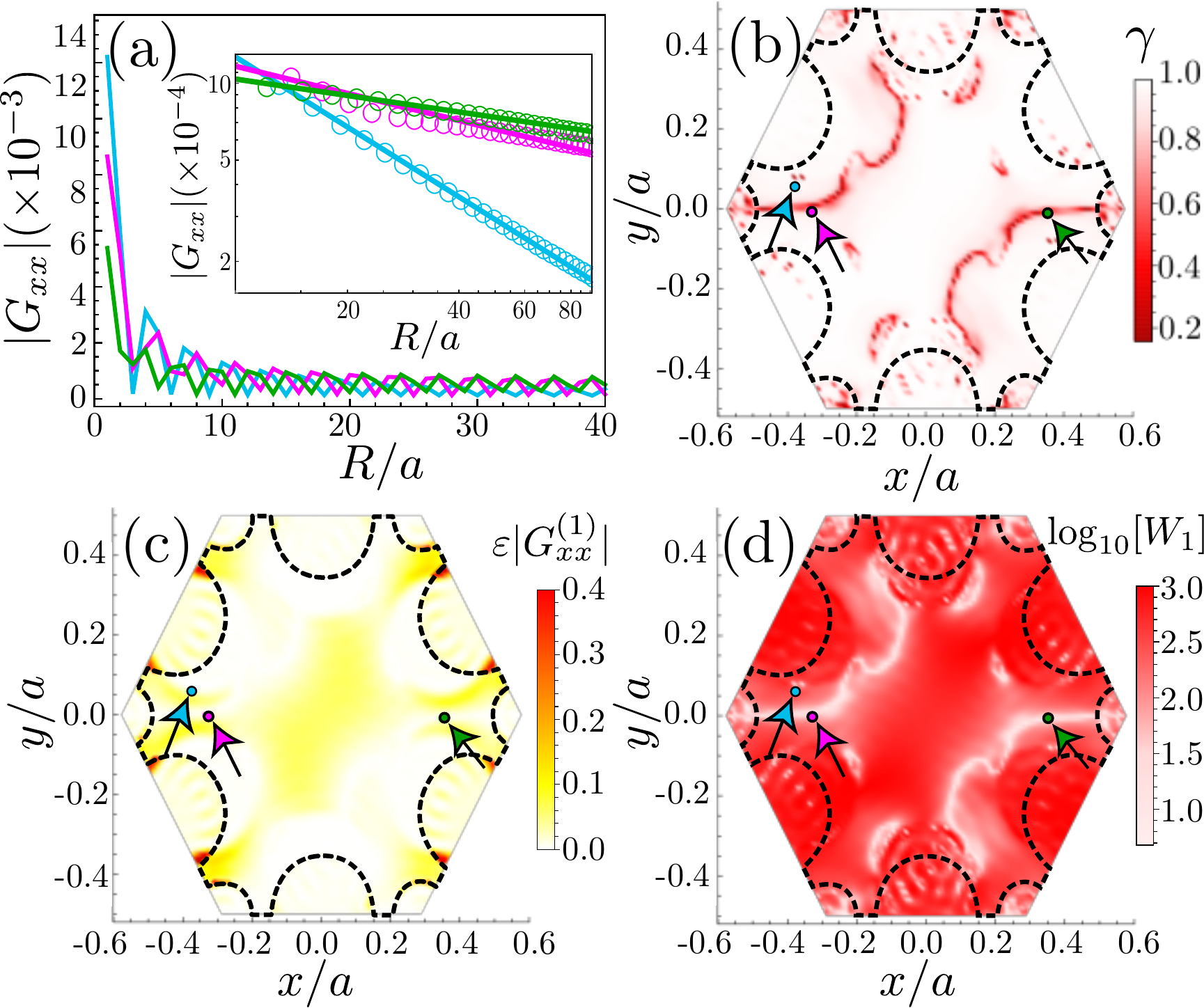}
   \caption{Behaviour of $G_{xx}$ for the direction of $\phi=\pi/6$ of a PhC structure with the same geometry parameters as in Fig.~\ref{fig: 2-bands-cone}. The transition wavelength of the emitter is fixed at $738$ nm such that the detuning is $\delta_A/2\pi=19.5$ GHz. (a) $|G_{xx}|$ as a function of $R$ for three different points in the unit cell that feature a different power-law, oscillatory, decay. The points of the unit cell chosen are depicted with arrows inside the density plots (b)-(d). The inset shows a selection of the maxima of the power-law decay plotted in logarithmic scale (in empty circles) together with a fitting to a power-law, $1/R^\gamma$, showing clearly how the different points feature different decay exponent $\gamma$. (b) Behaviour of the decay exponent ($\gamma$) obtained by a numerical fitting of $|G_{xx}|$ at all positions in the unit cell. (c) Behaviour of the strength of the interaction at the first neighbour position renormalized by the dielectric constant, $\varepsilon|G_{xx}^{(1)}|$, along the unit cell. (d) Ratio between the real and imaginary part of the Green function component at the first neighbour denoted by $W_1(\rr_1)$ and plotted in a logarithmic color scale.}
    \label{fig:Gxx}
\end{figure}

Let us start by considering the photon-mediated interactions of emitters with linearly polarized optical transitions oriented along one of the Cartesian components ($\hat{x}$, $\hat{y}$). As aforementioned, these interactions are given by the Green functions of Eq.~\eqref{eq:Gab}, that are expressed as sums of several decaying terms scaling with different power-laws. In previous works~\cite{Gonzalez-Tudela2018,Perczel2020a} the dominant contribution was shown to be given by a $1/r$-decay law. However, we find that when the position $\rr_1$ of the emitters is varied, the interference between the different terms can give rise to even longer-ranged interactions at certain points. This is observed in Fig.~\ref{fig:Gxx}(a), where we plot $|G_{xx}(\rr_1;\RR)|$ as a function of $R$ for a particular primitive lattice direction (other components and directions lead to qualitatively similar conclusions) for three different emitter positions in different colors, highlighted with arrows in Fig.~\ref{fig:Gxx}(b). At all emitter positions chosen, $|G_{xx}(\rr_1;\RR)|$ have an oscillatory and power-law decay behaviour. However, the decay range is faster in some positions with respect to the others. To make this more evident, we take only the maximum value of the oscillations within each period and plot them in the inset of Fig.~\ref{fig:Gxx}(a) in logarithmic scale, where one clearly observes that they follow a different power-law behaviour. 

To make a more detailed analysis of the change of this exponent we make a numerical fit of the envelopes of $|G_{xx}(\rr_1;\RR)|$ to a power-law $\propto 1/|\RR|^\gamma$ at all positions of the unit cell $\rr_1$ and plot the results of the fitting in Fig.~\ref{fig:Gxx}(b) in a color scale. In this figure, the white color denotes regions where the $1/r$ behaviour dominates ($\gamma=1$), whereas red denotes scalings with longer-range decays ($\gamma<1$). There, we observe how indeed most of the regions display the expected $1/r$ behaviour predicted in previous works. However, there  also appear other regions where the interference between the different terms lead to longer-ranged interactions. 

\begin{figure}[tb]
    \centering
    \includegraphics[scale=0.33]{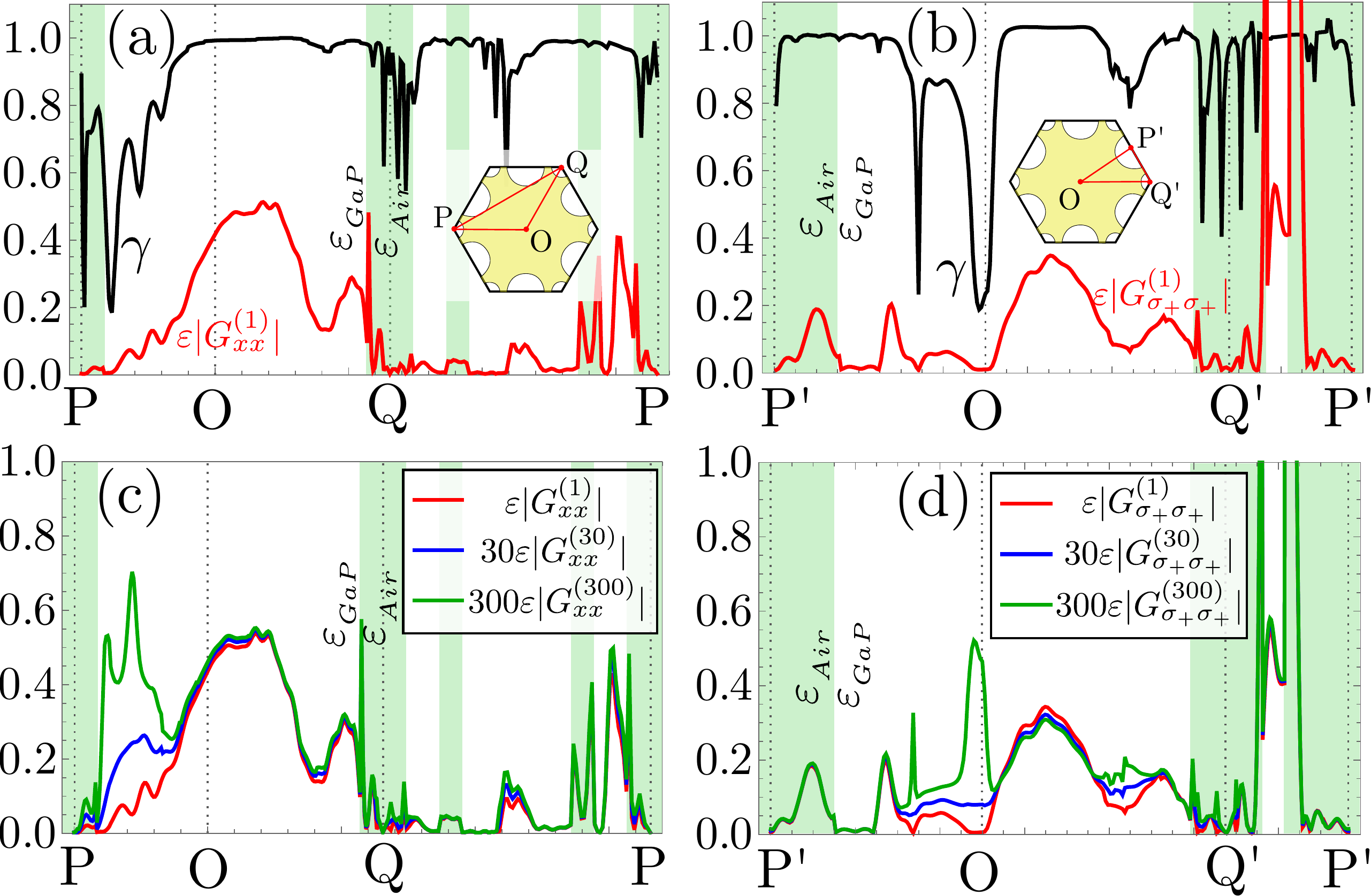}
    \caption{(a)-(b) The numerically obtained power-law exponent $\gamma$ and nearest-neighbour coupling strength are shown in panels (a) and (b) for linear ($|G^{(1)}_{xx}|$) and circularly polarized transitions ($|G^{(1)}_{\sigma_+\sigma_+}|$), respectively. The inset shows the path used to calculate the different curves in each polarization. (c)-(d) Comparison between the absolute value of the Green function at different distances for linear and circular polarization, respectively.  To appropriately compare between the first, 30th and 300th neighbour the factor $j$ is multiplying the value $\varepsilon |G_{xx(\sigma_+\sigma_+)}^{(j)}|$. Green shading represent regions along the path within the air holes.}
    \label{fig:tradeoff}
\end{figure}

Apart from the range of the interaction, another very relevant magnitude is their strength. This strength also depends on the emitter position $\rr_1$, since the mode function profile also changes along the unit cell. A way of characterizing this strength is by plotting the absolute value of the photon-mediated interactions between nearest-neighbouring atoms placed at different $\rr_1$ positions, $|G_{xx}(\rr,\rr+\mathbf{a}_1)|\equiv |G_{xx}^{(1)}(\rr)|$. This is what we plot in Fig.~\ref{fig:Gxx}(c), multiplying it by the dielectric index $\varepsilon$ so that the strength at the air/hole regions appear on a similar color scale. From this figure, we can make two important observations: first, the optimal position to couple the emitter is not at the center of the unit cell. For emitters within the dielectric, the largest coupling strength is obtained at the regions between the air holes. For emitters lying in the holes, the coupling is maximum at regions very close to the interface. However, trapping atoms close to surfaces is generally challenging due to Casimir-Polder forces~\cite{hung13a,Gonzalez-Tudela2015b}. Thus, in that case it would still be easier to place the atoms at the center of the holes, even if the coupling strength is smaller. The second important observation is that there is an apparent trade-off between the range and strength of interactions, since the red regions of panel (b) appears to be whiter (no coupling) in panel (c). We will explore this trade-off in more detail in Fig.~\ref{fig:tradeoff}.

Finally, let us note that in Figs.~\ref{fig:Gxx}(a-c) we plot the absolute value without differentiating the contributions from the real and imaginary part of the Green Function. However, as we introduced in the first part of the manuscript, both terms give rise to very different quantum dynamics: the real part leads to purely coherent exchanges ($J_{ij}$), whereas the imaginary part ($\Gamma_{ij}$) yields super/sub-radiant effects. Thus, in Fig.~\ref{fig:Gxx}(d) we characterize which term dominates by plotting the ratio between the real and imaginary part of the nearest-neighbour position:
\begin{equation}
W_1(\rr_1)=\left|\frac{\mathrm{Re}\left[G_{xx}^{(1)}(\rr_1)\right]}{\mathrm{Im}\left[G_{xx}^{(1)}(\rr_1)\right]}\right|\,.
\end{equation}

As shown in the legend accompanying the panel we use a logarithmic colorscale for $W_1(\rr)$ where the red (blue) color denotes $W_1(\rr)> (<) 1$, while white denotes the transition where $W_1(\rr)=1$. From the figure we can clearly observe how linearly polarized emitters lead to decoherence-free photon-mediated interactions ($W_1(\rr)> 1$) along the unit cell. This is in accordance to the results of the literature~\cite{Gonzalez-Tudela2018,Perczel2020a} and it is expected because of the vanishing density of states of the photonic bath around the Dirac point, i.e, $D(E)\propto |E|$.

\begin{figure}[tb]
    \centering
    \includegraphics[scale=0.48]{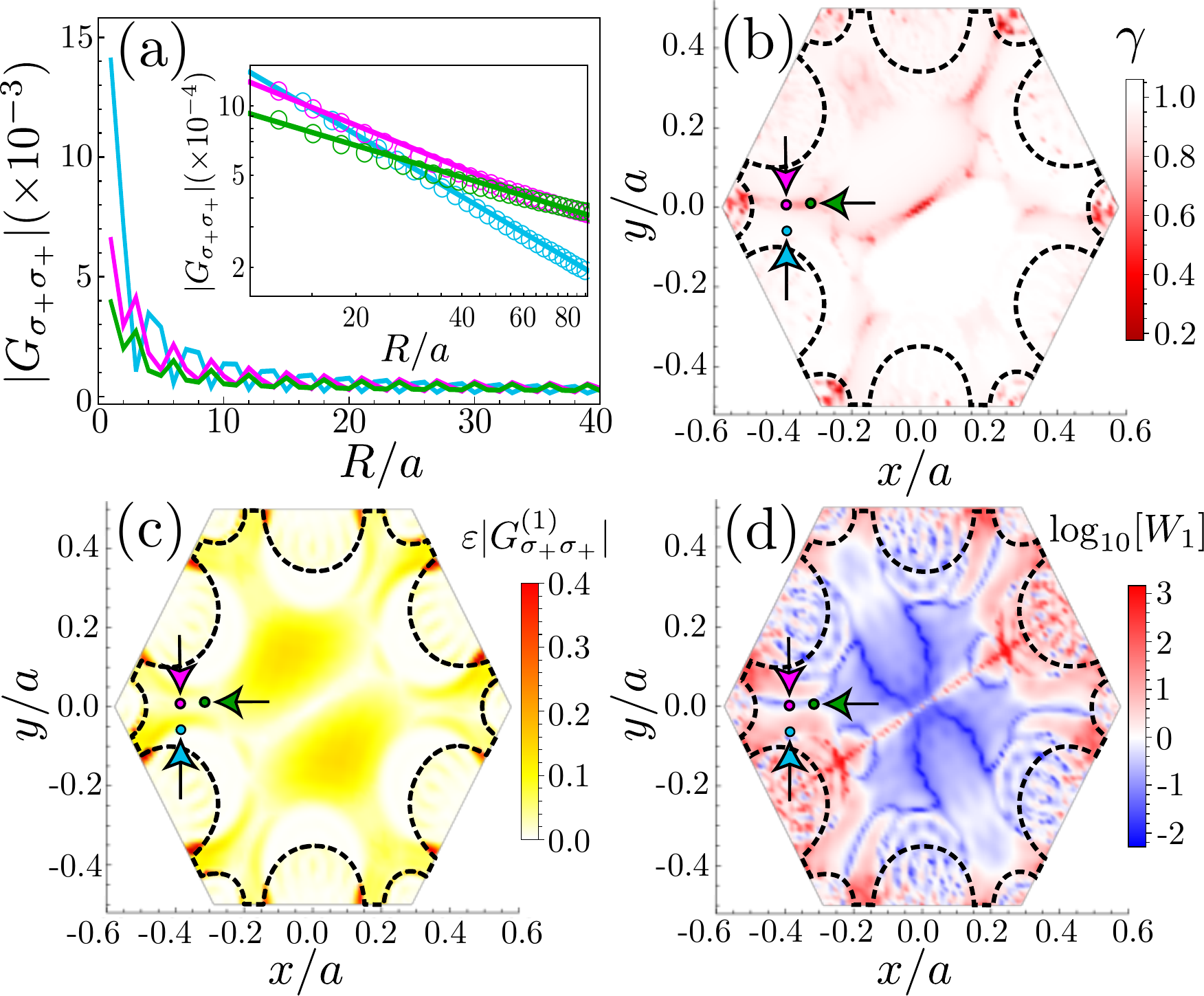}
    \caption{Behaviour of $G_{\sigma_+\sigma_+}$ for the direction of $\phi=\pi/6$ for the direction of $\phi=\pi/6$ for the same parameters than Fig.~\ref{fig:Gxx}. (a) $|G_{\sigma_+\sigma_+}|$ as a function of $R$ for three different points in the unit cell that feature a different power-law, oscillatory, decay. The points of the unit cell chosen are depicted with arrows inside the density plots (b)-(d). The inset shows a selection of the maxima of the power-law decay plotted in logarithmic scale (in empty circles) together with a fitting to a power-law, $1/R^\gamma$, showing clearly how the different points feature different decay exponent $\gamma$. (b) Behaviour of the decay exponent ($\gamma$) obtained by a numerical fitting of $|G_{\sigma_+\sigma_+}|$ at all positions in the unit cell. (c) Behaviour of the strength of the interaction at the first neighbour position renormalized by the dielectric constant, $\varepsilon|G^{(1)}_{\sigma_+\sigma_+}|$, along the unit cell. (d) Ratio between the real and imaginary part of the Green function component at the first neighbour denoted by $W_1(\rr_1)$ and plotted in a logarithmic color scale.}
    \label{fig:Gspsp}
\end{figure}

Let us now repeat this analysis with emitters with circularly polarized transitions. For the sake of illustration we focus again on a particular component of the Green Function, i.e., $G_{\sigma_+\sigma_+}$, although the conclusions can be extrapolated to the other components. We start by plotting again $|G_{\sigma_+\sigma_+}(\rr_1,\rr_1+\RR)|$ in Fig.~\ref{fig:Gspsp}(a) at several representative positions $\rr_1$, with different power-law scaling. This justifies again making the numerical fit to a power-law $\propto 1/|\RR|^\gamma$ at all unit cell positions, whose results we plot in Fig.~\ref{fig:Gspsp}(b). There, we observe how although the dominant exponent is the expected $\gamma=1$, there are several regions with longer-ranged interactions, as it occurs for the linearly polarized transition. Then, in Fig.~\ref{fig:Gspsp}(c) we plot the first-neighbour coupling strength multiplied by the permittiviy of the material, i.e., $\varepsilon |G^1_{\sigma_+\sigma_+}(\rr_1)|$. Differently from the linearly polarized case, the center of the unit cell is one of the least efficient places to couple the emitter. Again, the more favorable regions are the dielectric parts within the holes, although there also appear other places around the central part of the unit cell. At the holes, the pattern is very similar to the one of linear polarizations of Fig.~\ref{fig:Gxx}(c). The main difference with respect to the linearly polarized emitters is the behaviour of the real over imaginary part ratio,  $W_1(\rr)$, defined now for $G^{(1)}_{\sigma_+\sigma_+}(\rr_1)$, which we plot in Fig.~\ref{fig:Gspsp}(d). There, we observe how the photon-mediated interactions in most of the unit cell are dominated by the collective dissipative terms (blue regions), differently from the linearly polarized transitions (see Fig.~\ref{fig:Gxx}(d)). This is a consequence that for certain atomic positions of this structure $\mathrm{Re}\left[G_{xy}\right]\neq \mathrm{Re}\left[G_{yx}\right]$, which is known to lead to decay in circularly polarized transitions~\cite{Tang2010OpticalMatter,Garcia-Etxarri2013Surface-enhancedNanoantennas,Yoo2015ChiralResonators,Neuman2020NanomagnonicInteractions}. This change of the coherent nature of the interactions with polarization is something that was not captured in previous analysis because they either neglected polarization effects~\cite{Gonzalez-Tudela2018} or placed the emitters in positions where $G_{\sigma_+\sigma_+}\approx 0$\cite{Perczel2020a}.  

One important result from the previous analyses is the possibility of increasing interaction range (decreasing $\gamma$) by placing the emitter at different positions (see panel (b) of Figs.~\ref{fig:Gxx},~\ref{fig:Gspsp}). However, as we also see in panels (c) of the same figures, this increase comes at the price of reducing the absolute coupling strength. Since the color scales does not let us appreciate in detail this trade-off, in Figs.~\ref{fig:tradeoff}(a-b) we plot both $\gamma$ (black) and $\varepsilon |G_{\alpha\beta}^{(1)}(\rr)|$ (red) for the linear and circular polarization, respectively, and following a particular path of the unit cell depicted in the insets of this panels. There, we notice that indeed the minima of $\gamma$ coincide with positions where $|G_{\alpha\beta}^{(1)}(\rr)|\approx 0$. Thus, one natural question arises: does the increase of interaction ranges compensates the decrease of coupling strength to couple emitters at long distances? To answer this question in Fig.~\ref{fig:tradeoff}(c) and (d), we use the fit of the envelope power-law for plotting a comparison of the overall coupling strength at the first (red), 30th (blue), and 300th neighbour (green), multiplying the latter by a factor so that they appear in the same scale. There, we observe in spite of this trade-off there are regions within this path where the regions with $\gamma<1$ leads to an overall better coupling at long distances than the one of $\gamma=1$.

\section{Conclusions and outlook \label{sec:conclusions}}

Summing up, we have developed a semi-analytical theory based on the Guided-Mode Expansion and $\kp$ method to calculate the photon-mediated interactions in Dirac light-matter interfaces. An important advantage of our theory with respect to existing ones is that it enables us to calculate such interactions when emitters are placed at all positions of the unit cell. We benchmark our theory on a particular photonic structure, and find that it is possible to tune the interaction range of the emergent interactions by placing the emitter at different positions of the unit cell. Besides, we find the position that optimize the interactions between emitter at short and long-distances. Thus, we believe our theory and results can become a useful guide for future experimental designs of such Dirac light-matter interfaces.

An interesting outlook of our theory is applying it to other non-trivial band-structure points appearing in two-dimensional PhC slabs, such as Van-Hove singularities~\cite{galve17a,Gonzalez-Tudela2017b,Gonzalez-Tudela2017a,Yu2019}, where strong super/sub-radiant effects~\cite{Gonzalez-Tudela2017b,Gonzalez-Tudela2017a} or highly anisotropic coherent interactions~\cite{gonzaleztudela18f} have been predicted. Another research direction consists in harnessing the long-range nature of the photon-mediated in such Dirac light-matter interfaces for some of the quantum information and simulation applications mentioned in the introduction~\cite{shahmoon13a,eldredge17a,Kuwahara2020,Tran2020c,Tran2021b,hauke13a,richerme14a,gong16b,maghrebi16a,koffel12a,vodola14a,kastner11a}.

\begin{acknowledgements}
The authors acknowledge   support from i-COOP program from CSIC with project reference COOPA20280. AGT acknowledges  support   from   CSIC Research   Platform  on   Quantum   Technologies   PTI-001  and  from  Spanish  project  PGC2018-094792-B-100(MCIU/AEI/FEDER, EU). EPNB thanks financial support from the ``Programa de becas de excelencia doctoral del bicentenario - MINCIENCIAS 2019". HVP gratefully acknowledges funding by COLCIENCIAS under the project “Impact of phonon-assisted cavity feeding process on the effective light-matter coupling in quantum electrodynamics”, HERMES 47149. AGT acknowledge discussions with P. A. Huidobro about the decay of circularly polarized transitions.
\end{acknowledgements}


\newpage
\begin{widetext}

\begin{center}
\textbf{\large Supplemental Material: Photon-mediated interactions near a Dirac photonic crystal slab \\}
\end{center}
\setcounter{equation}{0}
\setcounter{figure}{0}
\makeatletter

\renewcommand{\thefigure}{SM\arabic{figure}}
\renewcommand{\thesection}{SM\arabic{section}}  
\renewcommand{\theequation}{SM\arabic{equation}}  

In this Supporting Information, we provide more details on: i) the Guided-Mode Expansion method to calculate the eigenfrequencies and eigenmodes of the Dirac photonic structure; ii) the $\kp$ method to approximate the eigenfrequencies and eigenmodes around the Dirac cones; iii) the derivation of the Green tensor by analytical integration using the expressions obtained by the $\kp$ method; iv) additional results for other components of the Green function.

\subsection{Guided-Mode Expansion method}
Time-independent electromagnetic waves satisfy the following eigenvalue problem
\begin{equation}
  \mathcal{L}\HH(\rr,z)= \nabla\times\left[\frac{1}{\varepsilon(\rr,z)}\nabla\times \HH(\rr,z)\right]=\frac{\omega^2}{c^2} \HH(\rr,z)\, ,
  \label{SI-Ec:1-MaxwellOp-t_Indep}
\end{equation}
where the eigenvalues are $\frac{\omega^2}{c^2}$, and the eigenfunctions are the magnetic field $\HH(\rr,z)$. Additionally, in materials with discrete symmetry translation, like photonic crystals (PhC), the eigenfunctions fulfill the following Bloch condition
\begin{eqnarray}
\HH_{ \kk}(\rr,z)&=&e^{i \kk\cdot\rr}\boldsymbol{u}_{ \kk}(\rr,z)\, ,
\label{SI-Ec:2-Bloch theorem}
\end{eqnarray}
where $\kk$ is the pseudo-momentum, and $\boldsymbol{u}_{ \kk}(\rr,z)$ is a periodic vector function with the same periodicity of the material, containing the real space dependence of the eigenfunctions within the unit cell.
\begin{figure}
    \centering
    \includegraphics[scale=0.7]{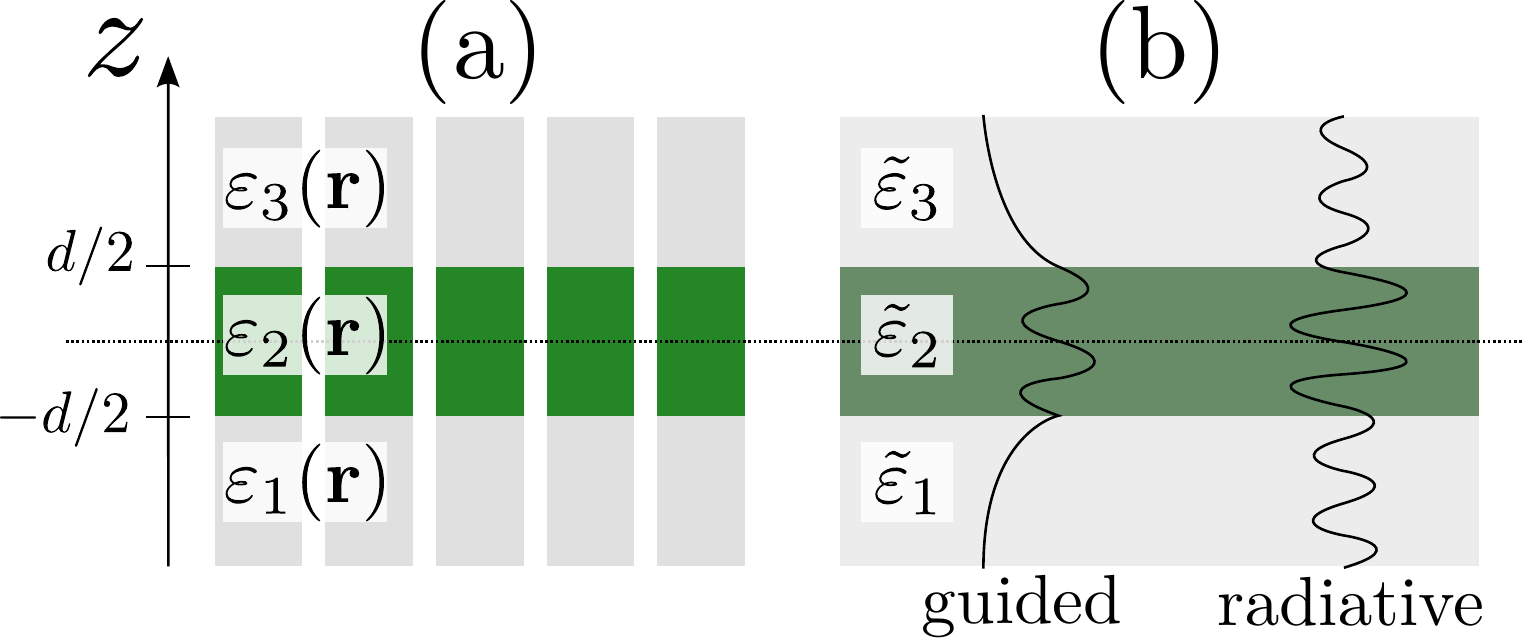}
    \caption{(a) Schematic picture a photonic crystal slab with periodic dependence of the permittivity, $\varepsilon_j(\rr)$. One can always separate the slab in three different regions according to the $z$-coordinate. (b) Schematic representation of the homogeneous slab problem which is the starting point of the GME method: it replaces the periodic structure of each of the regions of the photonic crystal slab of panel (a) by an effective permittivity, $\tilde{\varepsilon}_j$, in each spacial region. The homogeneous slab supports two types of modes: the guided mode, that are oscillating in the region 2 and evanescent in regions 1 and 3; and the radiative modes, that are oscillating in the three regions.}
    \label{SI-fig:Scheme-GME}
\end{figure}

Photonic crystal slabs can always be schematically represented as in Fig.~\ref{SI-fig:Scheme-GME}(a): they are structures with three different periodic materials in each region: $z<-d/2$, $-d/2\leq z\leq d/2$ and $z>d/2$. To solve Eq.~\eqref{SI-Ec:1-MaxwellOp-t_Indep} in these structures, one can use the Guided-Mode Expansion (GME) method \cite{andreani06a}, which initially considers a homogeneous slab waveguide with an effective permittivity given by the mean value of the permitivity $\tilde{\varepsilon}_j$ in the $xy$-plane for each region in $z$, as shown in Fig.~\ref{SI-fig:Scheme-GME}(b). To ensure the possibility of total internal reflection inside the region 2, the structures have to satisfy the condition $\tilde{\varepsilon}_{2}>\tilde{\varepsilon}_{1(3)}$. This homogeneous problem has two types of solutions: the guided modes, that are quantized and evanescent in the z direction at the regions $|z|>d/2$; and the radiative ones, that are a set of continuous modes and oscillating in z direction. The GME method use the former, plus Bloch theorem, to compose a new basis for solving the complete PhC slab problem. The basis can be separated into two orthogonal families, transverse electric (TE) and transverse magnetic (TM) modes. The TE modes can be written as:
\begin{equation}
\boldsymbol{h}^{(\text{TE})}_{\ggg,\mu}(\rr,z)=\frac{e^{i \ggg\cdot\rr}}{\sqrt{S}} \begin{cases}A_{\ggg,\mu}^{(3)}(\chi_{\mathbf{g},\mu}^{(3)}\hat{\ggg}+i \text{g} \hat{\mathbf{z}} )e^{-\chi_{\mathbf{g},\mu}^{(3)}(z-d/2)},& z>d/2\\
A_{\ggg,\mu}^{(2)}(-i q_{\mathbf{g},\mu}\hat{\ggg}+i \text{g} \hat{\mathbf{z}})e^{i q_{\mathbf{g},\mu}z}+B_{\ggg,\mu}^{(2)}(i q_{\mathbf{g},\mu}\hat{\ggg}+i \text{g} \hat{\mathbf{z}})e^{-i q_{\mathbf{g},\mu}z},& -d/2\leq z\leq d/2\\
B_{\ggg,\mu}^{(1)}(-\chi_{\mathbf{g},\mu}^{(1)}\hat{\ggg}+i \text{g} \hat{\mathbf{z}} )e^{\chi_{\mathbf{g},\mu}^{(1)}(z+d/2)},& z\leq -d/2\end{cases}
\end{equation}
and TM modes:
\begin{equation}
\boldsymbol{h}^{(\text{TM}}_{\ggg,\mu}(\rr,z)=\frac{e^{i \ggg\cdot\rr}}{\sqrt{S}}\hat{\mathbf{e}}_{\ggg} \begin{cases}C_{\ggg,\mu}^{(3)}e^{-\chi_{\ggg,\mu}^{(3)}(z-d/2)},& z>d/2\\
C_{\ggg,\mu}^{(2)}e^{i q_{\ggg,\mu}z}+D_{\ggg,\mu}^{(2)}e^{-i q_{\ggg,\mu}z},& -d/2\leq z\leq d/2\\
D_{\ggg,\mu}^{(1)}e^{\chi_{\ggg,\mu}^{(1)}(z+d/2)},& z\leq -d/2
\end{cases}
\end{equation}
where $A_{\ggg,\mu}^{(j)}$, $B_{\ggg,\mu}^{(j)}$, $C_{\ggg,\mu}^{(j)}$ and $D_{\ggg,\mu}^{(j)}$ are coefficients determined by the boundary conditions between the three spatial regions in the $z$-direction and by a normalization condition~\cite{andreani06a}; $\ggg$ is the projection of the wavevector in the $xy$-plane, $\hat{\mathbf{z}}$ is the unitary vector in $z$-direction, $S=\sqrt{3}a^2/2$ is the area of the unit cell, $\text{g}=|\ggg|$, $\hat\ggg=\ggg/\text{g}$, and $\hat{\mathbf{e}}_{\ggg}=\hat{\mathbf{z}}\times\hat\ggg$. The constants $\chi_{\mathbf{g},\mu}^{(j)}$ and $q_{\mathbf{g},\mu}$ are related with $\ggg$ and the frequencies modes by
\begin{eqnarray}
    \chi_{\ggg,\mu}^{(j)}&=&\sqrt{\text{g}^2 - \tilde{\varepsilon}_{j}\frac{\omega_{\text{g},\mu}^{2}}{c^2}}\, ,\\
    q_{\ggg,\mu}&=&\sqrt{\tilde{\varepsilon}_{2}\frac{\omega_{\textbf{g},\mu}^{2}}{c^2}-\text{g}^2}\, ,
\end{eqnarray}
The eigenfrequencies of the TE polarized modes are obtained by solving the following transcendental equation:
\begin{equation}
q_{\mathbf{g},\mu}(\chi_{\mathbf{g},\mu}^{(1)}+\chi_{\mathbf{g},\mu}^{(3)})\cos(q_{\mathbf{g},\mu}d)+(\chi_{\mathbf{g},\mu}^{(1)}\chi_{\mathbf{g},\mu}^{(1)}-q_{\mathbf{g},\mu}^2)\sin(q_{\mathbf{g},\mu}d)=0\, ,
\end{equation}
whereas the TM polarized frequencies are given by:
\begin{equation}
\frac{q_{\mathbf{g},\mu}}{\tilde{\varepsilon}_{2}}\left(\frac{\chi_{\mathbf{g},\mu}^{(1)}}{\tilde{\varepsilon}_{1}}+\frac{\chi_{\mathbf{g},\mu}^{(3)}}{\tilde{\varepsilon}_{3}}\right)\cos(q_{\mathbf{g},\mu}d)+\left(\frac{\chi_{\mathbf{g},\mu}^{(1)}\chi_{\mathbf{g},\mu}^{(3)}}{\tilde{\varepsilon}_{1}\tilde{\varepsilon}_{3}}-\frac{q_{\mathbf{g},\mu}^2}{{\tilde{\varepsilon}_{2}^{2}}}\right)\sin(q_{\mathbf{g},\mu}d)=0\, .
\end{equation}

For convenience in the derivation of $\kp$-approximation, we can introduce the guided periodic function, $\mathbf{U}_{\ggg,\mu}(\rr,z)$, of each homogeneous guided modes ($ \boldsymbol{h}_{\ggg,\mu}(\rr,z)$) as follows:
\begin{equation}
    \boldsymbol{h}_{\ggg,\mu}(\rr,z)=e^{i\kk\cdot\rr}\mathbf{U}_{\ggg,\mu}(\rr,z)\, \rightarrow \, \mathbf{U}_{\ggg,\mu}(\rr,z)=e^{-i\kk\cdot\rr}\boldsymbol{h}_{\ggg,\mu}(\rr,z)\, ,
\end{equation}

Using these guided modes for the homogeneous slab, we can expand the magnetic field of the PhC as
\begin{equation}
    \mathbf{H}_{\kk,n}(\rr,z)=\sum_{\GG,\mu} c_n(\kk+\GG,\mu) \boldsymbol{h}_{\kk+\GG,\mu}(\rr,z)\, ,
    \label{SI-Ec:3-GME H_Field}
\end{equation}
where we write $\ggg$ as the sum of $\kk$ a vector in the first Brillouin zone and $\GG$ a vector of the reciprocal lattice, such that we have the information about the periodicity of the structure already incorporated in the expansion. Here, the index $\mu$ runs over the different homogeneous guided modes, and we also introduce the index $n$ related to denote the different modes (bands) of the PhC slab. Then, using the expansion of $\HH_{\kk,n}$ of Eq.~\eqref{SI-Ec:3-GME H_Field} and the orthogonality of the guided modes, the eigenvalue Eq.~\eqref{SI-Ec:1-MaxwellOp-t_Indep} is rewritten as
\begin{equation}
    \sum_{\ggg,\nu}\mathcal{H}_{\ggg,\mu;\ggg',\nu} c_n(\ggg,\nu)=\left(\frac{\omega^{(n)}_{\kk}}{c}\right)^{2}c_n(\ggg,\mu)\, ,
    \label{SI-Ec:4-GME-MatrixEigenProblem}
\end{equation}
where the sum over $\ggg=\kk+\GG$ is equivalent to the sum over the lattice vector $\GG$ because $\kk$ is fixed for each eigenvalue calculation. The matrix elements $ \mathcal{H}_{\ggg,\mu;\ggg',\nu}$ read
\begin{equation}
    \mathcal{H}_{\ggg,\mu;\ggg',\nu}=\int_{V_c}\frac{1}{\varepsilon(\rr)}(\nabla\times\boldsymbol{h}^{*}_{\ggg,\mu}(\rr))\cdot(\nabla\times\boldsymbol{h}_{\ggg',\nu}(\rr))d\rr dz\, ,
\end{equation}
where $V_c$ is the integration volume corresponding to the unit cell in the $xy$-plane and $(-\infty,\infty)$ in the $z$-direction. An explicit expression for the matrix elements in terms of the guided modes can be found in Ref.~\citenum{andreani06a}. Eq.~\eqref{SI-Ec:4-GME-MatrixEigenProblem} is, in principle, an infinite linear matrix eigenproblem that must be truncated to be solved numerically. Solving this truncated eigenproblem, one can calculate the eigenfrequencies $\omega_\kk^{(n)}$ and the eigenvectors containing the coefficients $c_n(\kk+\GG,\mu)$ that expand the magnetic field of the PhC modes. Due to the reflection symmetry at the horizontal plane at $z=0$ of our structure ($z\rightarrow -z$), it is possible to classify the solutions into two families, odd (TM-like) and even (TE-like) eigenmodes. We are interested in the TE-like modes because they have non zero electric fields at $z=0$ where we put the emitters, so we only take into account the modes with even symmetry under reflection in $z$.

\subsection{$\kp$ method for Dirac cones}

In the main text, the $\kp$ method is used to obtain analytical expressions of $\omega_{\kk}^{(n)}$, $\HH_{\kk,n}(\rr,z)$ and $\EE_{\kk,n}(\rr,z)$ for momenta $\kk$ around the Dirac points $\KK^{(')}$. Here, we provide more details about this procedure.

According to Bloch theorem, the solution to the magnetic field at the $\KK$ point has the form of $\HH_{ \KK}(\rr)=e^{i \KK\cdot\rr}\boldsymbol{u}_{ \KK}(\rr)$. Using the GME method, we have an expression of the periodic function $\lbrace \boldsymbol{u}_{\KK,n}\rbrace$ which reads:
\begin{equation}
    \boldsymbol{u}_{\KK,n}(\rr,z)=\sum_{\GG,\mu}c_n(\KK+\GG,\mu)\mathbf{U}_{\KK+\GG,\mu}(\rr,z)\, ,
    \label{SI-Ec:PeriodicBasis}
\end{equation}

We use the set $\lbrace \boldsymbol{u}_{\KK,n}\rbrace$ as a basis to expand $\boldsymbol{u}_{\kk,n}$ for $\kk=\KK+\qq$, with which the magnetic field can be written as:
\begin{equation}
\HH_{\kk,n}(\rr,z)=e^{i \kk\cdot\rr}\boldsymbol{u}_{\kk,n}(\rr,z)=e^{i \qq\cdot\rr} e^{i \KK\cdot\rr}\sum_{j}C_{j}\boldsymbol{u}_{\KK,j}(\rr,z)\, , \label{SI-Ec:6-k.P MagneticField}
\end{equation}

By putting this magnetic field in Eq.~\eqref{SI-Ec:1-MaxwellOp-t_Indep}, we obtain the following eigenproblem~\cite{Ming1994}
\begin{equation}
    \sum_{j}\mathsf{H}_{l,j} C_{j}=\frac{\omega_{\kk,n}^{2}}{c^{2}}C_{l}\, ,
    \label{SI-Ec:7-k.P-general_eigenproblem}
\end{equation}
where, 
\begin{equation}
    \mathsf{H}_{l,j}=\frac{\omega^2_{\KK,j}}{c^{2}}\delta_{l,j}+ \qq\cdot\left[-i\boldsymbol{p}_{l,j}+i \boldsymbol{p}_{j,l}^{*}+2q_{lj}\KK-\mathbb{Q}_{l,j}\cdot\KK-\mathbb{Q}^{T}_{l,j}\cdot\KK\right]-\qq\cdot\mathbb{Q}_{l,j}\cdot\qq+q_{l,j} \qq\cdot\qq
    \label{SI-Ec:8-k.p-MatrixElements}
\end{equation}
with
\begin{align}
    \boldsymbol{p}_{l,j}&= \int_{V_{c}}d\rr dz\; \boldsymbol{u}^{*}_{\KK,l}(\rr,z)\times\left[\frac{\nabla\times\boldsymbol{u}_{\KK,j}(\rr,z)}{\varepsilon(\rr,z)}\right]\\
    \mathbb{Q}_{l,j}&= \int_{V_{c}}d\rr dz\; \frac{\boldsymbol{u}^{*}_{\KK,l}(\rr,z)\boldsymbol{u}_{\KK,j}(\rr,z)}{\varepsilon(\rr,z)}\\
    q_{l,j}&=\text{Tr}(\mathbb{Q}_{l,j})
\end{align}

The $\mathsf{H}_{l,j}$ matrix elements have in general terms of order $\qq^0$, $\qq^1$ and $\qq^2$, such that solving Eq.~\eqref{SI-Ec:7-k.P-general_eigenproblem} yields a perturbative solution of the eigenvalues and eigenmodes. To solve this equation numerically. we use a finite set of $\boldsymbol{u}_{\KK,j}(\rr)$, and obtain the $\lbrace C_j\rbrace$ coefficients. With them, we can reconstruct the magnetic field of the PhC slab using Eq.~\eqref{SI-Ec:6-k.P MagneticField}, and finally, obtain the corresponding electric field using Maxwell equations.

For the particular Dirac-cone scenario, this general perturbative eigenvalue problem can be further simplified. First, because for Dirac cones around the $\KK^{(')}$ points, we have only two degenerate eigenvalues at the two first bands. Thus, we only need to consider two orthonormal modes from this eigenvalue to solve a perturbative degenerate problem. Second, because the dispersion at these symmetric points is approximately linear, such that we only include the terms of $\mathsf{H}_{l,j}$ with zero and first order in $\qq$. With these two considerations, Eq.~\eqref{SI-Ec:7-k.P-general_eigenproblem} is simplified to

\begin{equation}
    \sum_{j=1}^{2}\qq\cdot\PP_{l,j} C_{j}=\frac{\omega_{\kk,n}^{2}-\omega^2_{\KK,n}}{c^{2}}C_{l}\,,
    \label{SI-Ec:9-k.P-eigenproblem}
\end{equation}
where $\PP_{l,j}=i(-\boldsymbol{p}_{l,j}+\boldsymbol{p}_{j,l}^{*})+2q_{lj}\KK-(\mathbb{Q}_{l,j}+\mathbb{Q}^{T}_{l,j})\cdot\KK$. Besides, as a consequence of the \emph{deterministic degeneracy}~\cite{Mei2012}, there are particular relationships for the $\PP_{l,j}$'s, that are, $\PP_{2,2}=-\PP_{1,1}$ and $\PP_{2,1}=\PP_{1,2}^{*}$, which are approximately satisfied in numerical calculations. In this way, Eq.~\eqref{SI-Ec:9-k.P-eigenproblem} has the following matrix form
\begin{equation}
\begin{pmatrix}
\qq\cdot\PP_{11}&& \qq\cdot\PP_{12}\\
\qq\cdot\PP^{*}_{12}&& -\qq\cdot\PP_{11}
\end{pmatrix}\begin{pmatrix}
\xi_{\pm}\\
\eta_{\pm}
\end{pmatrix}=\Delta\lambda_{\pm}\begin{pmatrix}
\xi_{\pm}\\
\eta_{\pm}
\end{pmatrix}
\end{equation}
where the eigenvalues are $\Delta\lambda_{\pm}=\pm\frac{\omega_{\kk\pm}^{2}-\omega^2_{0}}{c^{2}}$, and the subindices $\pm$ are related to the upper and lower bands of the cone. Then, solving the $2\times2$ matrix eigenproblem, we obtain $\Delta\lambda_{\pm}$ and $(\xi_{\pm},\eta_{\pm})$. The former gives us the correction to the squared frequencies in a linear way
\begin{equation}
    \omega^{2}_\pm(\KK^{(')}+\qq)\approx \omega_D^{2}\left(1\pm m|\qq|\right)  \,,\label{SI-Ec: wkSqaprox}
\end{equation}
where $m$ is the slope of the cone, which depends of the geometric parameters of the slab. Around the Dirac cones $m|\qq|\ll 1$, such that the square root of Eq.~\eqref{SI-Ec: wkSqaprox} can be approximated by:
\begin{equation}
    \omega_\pm(\KK^{(')}+\qq)\approx \omega_D\left(1\pm \frac{1}{2}m|\qq|\right)=\omega_D\pm v|\qq|  \,,\label{SI-Ec: wkaprox}
\end{equation}
where $v$ is the group velocity of the Dirac cone. Using this velocity, we define $m= 2 \frac{v}{c}\frac{\omega_D}{c}$ and the eigenvalues as $\Delta_\pm=\pm 2\frac{v}{c}\frac{\omega_D}{c}|\qq|$.

The coefficients $(\xi_\pm,\eta_\pm)$  give us the contribution of each orthonormal $\boldsymbol{u}_{\KK,j}(\rr)$ to the magnetic fields ($\HH_{\kk,n}(\rr)$). With these coefficients, the magnetic fields are written as follows
\begin{equation}
\HH_{\kk,\pm}(\rr)\approx e^{i \kk\cdot\rr}\left[\xi_{\pm}\boldsymbol{u}_{\KK,1}(\rr)+\eta_{\pm}\boldsymbol{u}_{\KK,2}(\rr)\right]
\end{equation}
Remembering that $\kk=\KK+\qq$ and $\HH_{\KK,j}(\rr,z)=e^{i\KK\cdot\rr}\boldsymbol{u}_{\KK,j}(\rr,z)$, we can finally express
\begin{equation}
\HH_{\kk,\pm}(\rr)\approx e^{i \qq\cdot\rr}\left[\xi_{\pm}\HH_{\KK,1}(\rr)+\eta_{\pm}\HH_{\KK,2}(\rr)\right]
\end{equation}
The electric field can be derived from the relation $\EE=\frac{i}{\varepsilon(\rr)}\frac{c}{\omega}\nabla\times\HH$, as
\begin{equation}
\EE_{\kk,\pm}(\rr)\approx e^{i \qq\cdot\rr}\left[\xi_{\pm}\EE_{\KK,1}(\rr)+\eta_{\pm}\EE_{\KK,2}(\rr)\right]\label{SI-Ec:Efield_with_kP}
\end{equation}
For Dirac cones, these coefficients can be expressed in trigonometric functions, as shown in the main text: 
\begin{align}
    \xi_+=&\sin\left[\frac{\phi_\qq-\delta_{\KK^{(')}}}{2}\right]\, ,\label{SI-Ec:Xi_plus}\\
    \eta_+=&\mp\cos\left[\frac{\phi_\qq-\delta_{\KK^{(')}}}{2}\right]\, ,\\
    \xi_-=&\pm\cos\left[\frac{\phi_\qq-\delta_{\KK^{(')}}}{2}\right]\, ,\\
    \eta_-=&\sin\left[\frac{\phi_\qq-\delta_{\KK^{(')}}}{2}\right]\, ,\label{SI-Ec:Eta_minus}
\end{align}
where the symbols $\pm$ and $\mp$ are for $\KK/\KK'$, $\tan(\phi_\qq)=q_y/q_x$, and the phases $\delta_{\KK^{(')}}=\pm\pi/6$. 

Additionally, we can use guided modes to express and calculate the vectors $\PP_{l,j}$ and the matrix elements $\mathsf{H}_{l,j}$. Using the expansion of $\boldsymbol{u}_{\KK,j}(\rr,z)$ in terms of the guided periodic functions $\mathbf{U}_{\ggg,j}(\rr,z)$, we obtain the following expressions.
\begin{align}
    \boldsymbol{p}_{l,j}&=\sum_{\ggg,\mu}\sum_{\ggg',\nu} c_l^{*}(\ggg',\nu)\PPP_{\ggg',\nu;\ggg,\mu} c_j(\ggg,\mu)\,,\\
    \mathbb{Q}_{l,j}&= \sum_{\ggg,\mu}\sum_{\ggg',\nu} c_l^{*}(\ggg',\nu)\mathcal{Q}_{\ggg',\nu;\ggg,\mu} c_j(\ggg,\mu)\,,
\end{align}
where,
\begin{align}
    \PPP_{\ggg',\nu;\ggg,\mu}=\int_{V_{c}}d\rr dz\; \mathbf{U}_{\ggg',\nu}^{*}(\rr,z)\times\left[\frac{\nabla\times\mathbf{U}_{\ggg,\mu}(\rr,z)}{\varepsilon(\rr,z)}\right]\,,\\
    \mathcal{Q}_{\ggg',\nu;\ggg,\mu}=\int_{V_{c}}d\rr dz \frac{\mathbf{U}^{*}_{\ggg',\nu}(\rr)\mathbf{U}_{\ggg,\mu}(\rr)}{\varepsilon(\rr)}\,.
\end{align}

The elements $\PPP_{\ggg',\nu;\ggg,\mu}$ and $\mathcal{Q}_{\ggg',\nu;\ggg,\mu}$ are obtained by integration of guided modes with TE or TM polarization, and each polarization case have different analytical expressions, so we add the superscripts (TE-TE), (TE-TM), (TM-TE) and (TM-TM), for example:
\begin{equation}
    \PPP_{\ggg',\nu;\ggg,\mu}^{\mathrm{(TE-TM)}}=\int_{V_{c}}d\rr dz\; \mathbf{U}^{*\mathrm{(TE)}}_{\ggg',\nu}(\rr,z)\times\left[\frac{\nabla\times\mathbf{U}^{\mathrm{(TM)}}_{\ggg,\mu}(\rr,z)}{\varepsilon(\rr,z)}\right]
\end{equation}

The explicit expressions for each element can also be analytically calculated but we do not write them here because they are so large that do not add any further physical insight.
\begin{figure}[tb]
    \centering
    \includegraphics[scale=0.5]{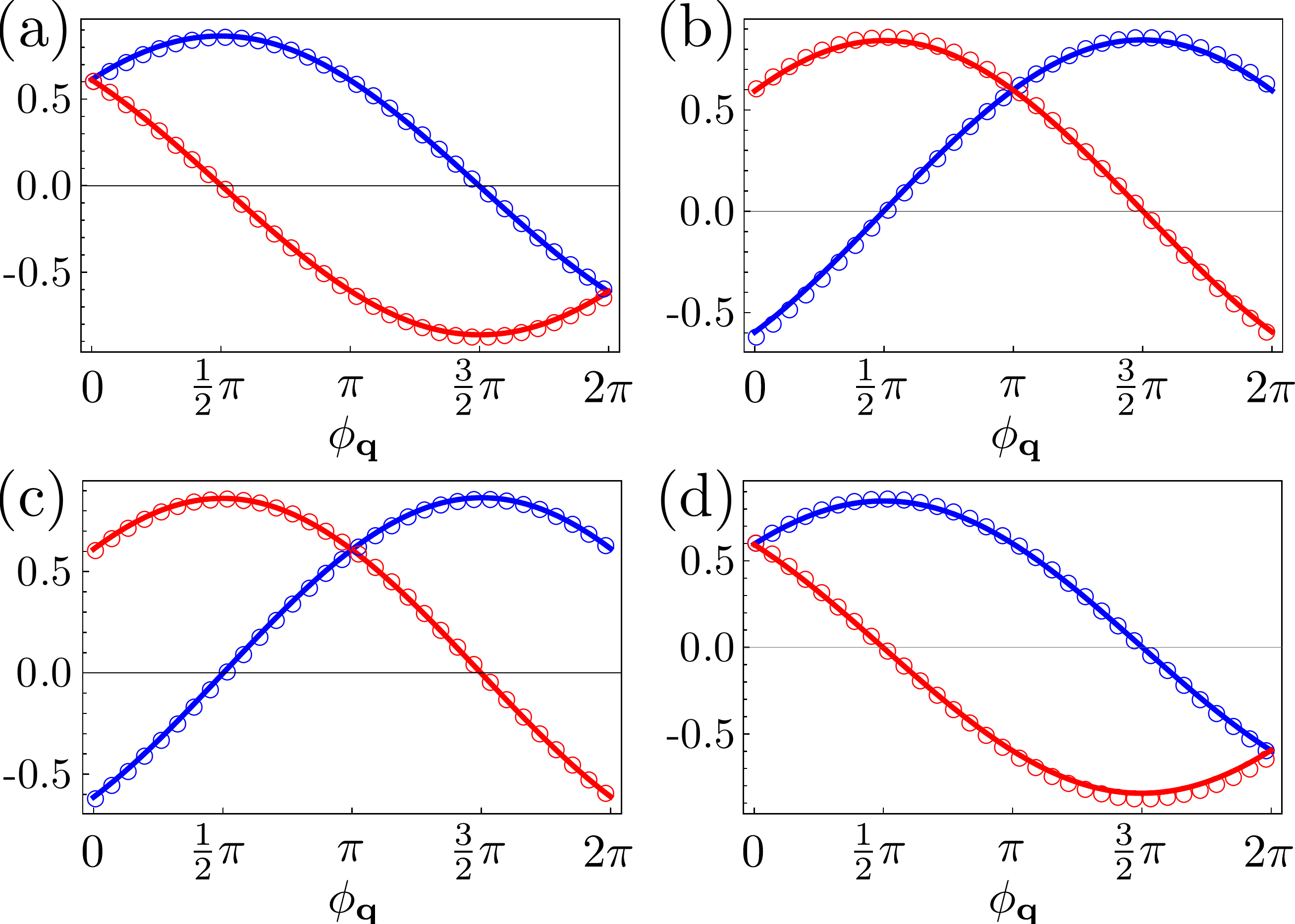}
    \caption{Comparison of the electric fields at $\rr=0$ obtained using the empirical ansatz of Ref.~\citenum{Perczel2020a} (empty circles) and the semi-analytical expressions obtained from the $\kp$-method (solid lines). The blue (red) color denote the x(y)-component of the electric fields. (a-b) [(c-d)] Electric field around $\KK$ [$\KK'$] for the lower and upper bands, respectively. The geometric parameters of the PhC slab are: $r_s=0.0833 a$, $r_l=0.15 a$,  $l=0.4 a$, $d=0.25a$. }
    \label{SI-fig:Lukin_comparison}
\end{figure}

To conclude this section, let us benchmark the approximated electric fields obtained through the $\kp$ expansion doing two sanity checks. First, let us compare our results (Eq.~\eqref{SI-Ec:Efield_with_kP}) with the ones derived by~\citeauthor{Perczel2020a} for the central position of the unit cell, which read:
\begin{equation}
\EE_{\kk,\pm}(\rr=0)\approx E_0\left[\sin\left(\frac{\phi_\qq}{2}\mp (\pm)\frac{\pi}{4}\right)\hat{\mathbf{x}}\pm (\mp)\sin\left(\frac{\phi_\qq}{2}\pm (\mp)\frac{\pi}{4}\right)\hat{\mathbf{y}}\right]\label{SI-Ec:Efield_Lukin_at_K}
\end{equation}
for $\kk$ around $\KK^{(')}$. Since the direct comparison between formulas is not obvious, in Fig.~\ref{SI-fig:Lukin_comparison} we plot the dependence of the electric field components with the angular variable $\phi_\qq$, calculated by $\kp$-method and by the ansatz of Eq.~\eqref{SI-Ec:Efield_Lukin_at_K}. To do it, we fix the position to $\rr=0$ and $|\qq|=0.03|\KK|$. We plot in blue (red) the x(y)-component of the electric field according to Ref.~\citenum{Perczel2020a} (in empty circles) or to $\kp$-approximated expressions (in solid-lines) showing a very good agreement for both the $\KK$ (panels (a-b)) and $\KK'$ (panels (c-d)) points. 

Apart from this comparison, it is important to ensure the capability of our approximation to capture the electric field dependence at other positions of the unit cell. A way to benchmark it consists in fixing a particular $\qq$ and calculate the field mode distribution along the whole unit cell using both a full numerical approach (GME) and the approximated expression. This is what we show in Fig.~\ref{SI-fig:GME_comparison}, where we compare in panels (a)-(b) [(c)-(c)] both approaches for $\kk$ close to the $\KK$ [$\KK'$] points.

\begin{figure}[tb]
    \centering
    \includegraphics[scale=0.5]{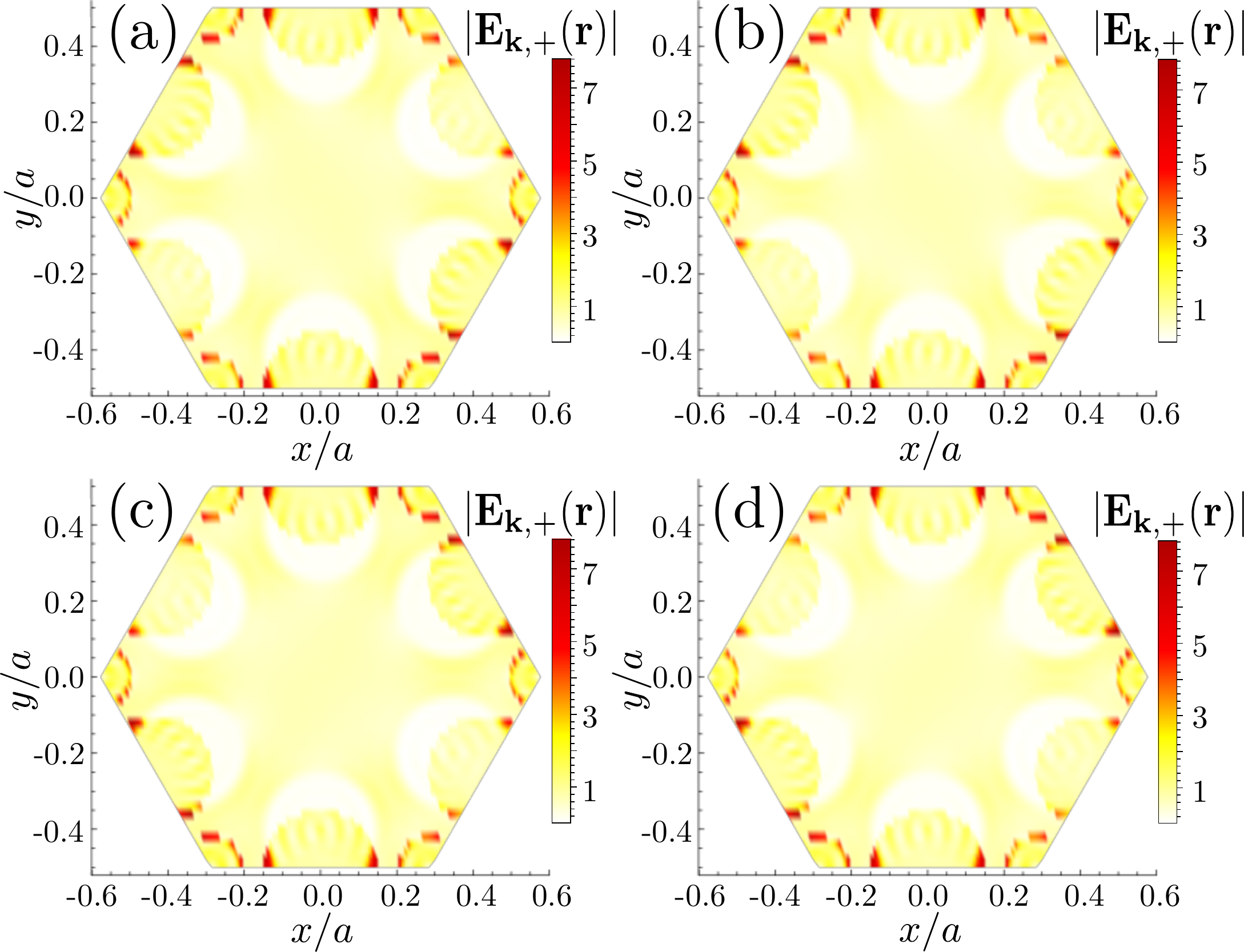}
    \caption{(a-b) [(c-d)] Electric fields in the whole unit cell for an eigenmode in the upper band with momentum $\kk=\KK [\KK']+\qq$, with  $|\qq|=0.03|\KK|$ and $\phi_\qq=0$. In the two panels we compare the results obtained numerically with GME method, panels (a) and (c), and the approximated fields obtained by the $\kp$ approximation, panels (b) and (d). The geometric parameters of the PhC slab are the same as in Fig.~\ref{SI-fig:Lukin_comparison}. }
    \label{SI-fig:GME_comparison}
\end{figure}

\subsection{Green functions: analytical integration using $\kp$ method}

The calculation of the two-point Green function is made using the definitions:
\begin{align}
G_{\alpha\beta}(\rr_1;\RR)=\iint_\mathrm{BZ} \frac{d^2\pp}{(2\pi)^2} g_{\alpha\beta}(\pp)\,,\label{SI-Ec:intGij}
\end{align}
and 
\begin{align}
g_{\alpha\beta}(\pp)=\frac{\sqrt{3}a^2}{2}c^2\sum_{n}\frac{\EE^{(n)*}_{\pp,\alpha}(\rr_1)\EE^{(n)}_{\pp,\beta}(\rr_1+\RR)}{\omega_A^2-\left(\omega^{(n)}_{\pp}\right)^2}
\end{align}
We are interested in emitters frequencies very close to the Dirac point, so we make the following approximations:
\begin{itemize}
    \item We only consider the contributions of the first two modes ($n=1,2$), that are, the two bands that compose the Dirac cone.
    
    \item We only consider $\kk$'s close to the high symmetry points $\KK^{(')}$; these $\kk$'s are delimited by a circular cutoff around  $\KK^{(')}$, i. e., $|\kk-\KK^{(')}|\leq q_c$. In this region, the dispersion relation is linear with $\qq=\kk-\KK$, as the $\kp$ method shows, and the electrical field for the two bands are given by Eq.~\eqref{SI-Ec:Efield_with_kP}. 
    
    \item As $\omega_A\approx \omega_D$, we consider that
    $$\omega_A^{2}-(\omega_\pp^{(n)})^2\approx 2\omega_A (\omega_A-\omega_\pp^{(n)}) $$
    
    \item We are interested in points separated by lattice displacements, i.e., $\rr_2=\rr_1+\RR$, so the fields at the two points are related by the Bloch theorem as follows:
    $$\EE_{\kk,\pm}(\rr_2)=e^{i \kk\cdot\RR}\EE_{\kk,\pm}(\rr_1)$$
\end{itemize}

Using these approximations, the integral of Eq.~\eqref{SI-Ec:intGij} are written as
\begin{equation}
  G_{\alpha,\beta}(\rr_1,\rr_2,\omega_A)=\frac{\sqrt{3} a^2}{2}\frac{c^{2}}{(2\pi)^2}\left[ I_{\alpha,\beta}^{(\KK,+)}+I_{\alpha,\beta}^{(\KK,-)}+I_{\alpha,\beta}^{(\KK',+)}+I_{\alpha,\beta}^{(\KK',-)}\right]
\end{equation}
where,
\begin{align}
    I_{\alpha,\beta}^{(\KK,\pm)}=& \frac{1}{2\omega_A}\int_{\KK}q dqd\phi_{\qq} \frac{E^{(\KK,\pm)*}_{\qq,\alpha}(\rr_{1})E^{(\KK,\pm)}_{\qq,\beta}(\rr_{2})}{\omega_{A}-\omega_{D}\mp v q}\\
    I_{\alpha,\beta}^{(\KK',\pm)}=& \frac{1}{2\omega_A}\int_{\KK'}q dqd\phi_{\qq} \frac{E^{(\KK',\pm)*}_{\qq,\alpha}(\rr_{1})E^{(\KK',\pm)}_{\qq,\beta}(\rr_{2})}{\omega_{A}-\omega_{D}\mp v q}
\end{align}
the subindex $\KK^{(')}$ in the integral denotes that the integrals are made around the Dirac point at $\KK^{(')}$. Here, we have considered a change of the integration variable $\pp=\KK^{(')}+\qq$ and written $\qq=(q\cos(\phi_\qq),q\sin(\phi_\qq))$ in polar coordinates. The limits for the integration variables are $0\leq q\leq q_c$ and $0\leq \phi_\qq \leq 2\pi$.

Now, we use the explicit form of the electrical field as found in Eqs.~\eqref{SI-Ec:Efield_with_kP}-\eqref{SI-Ec:Eta_minus} to write the following product of the components:
\begin{align}
    E^{(\KK,\pm)*}_{\qq,\alpha}(\rr_{1})E^{(\KK,\pm)}_{\qq,\beta}(\rr_{2})=&e^{i \KK\cdot \RR} e^{i \qq\cdot \RR}\left\lbrace\xi_\pm^2 E_{\KK,1,\alpha}^{*}(\rr_1)E_{\KK,1,\beta}(\rr_1)+\eta_\pm^2 E_{\KK,2,\alpha}^{*}(\rr_1)E_{\KK,2,\beta}(\rr_1)\right.\nonumber\\
    &\left.+\xi_\pm\eta_\pm\left[E_{\KK,1,\alpha}^{*}(\rr_1)E_{\KK,2,\beta}(\rr_1)+E_{\KK,2,\alpha}^{*}(\rr_1)E_{\KK,1,\beta}(\rr_1)\right]\right\rbrace
\end{align}

Considering the trigonometric expression for the coefficients $\xi_\pm$ and $\eta_\pm$ (Eqs.~\eqref{SI-Ec:Xi_plus}-\eqref{SI-Ec:Eta_minus}), we have three types of terms:
\begin{align}
    &\sin^{2}\left(\frac{\phi_\qq-\delta_\KK}{2}\right)=\frac{1-\cos(\phi_\qq-\delta_\KK)}{2}\\
    &\cos^{2}\left(\frac{\phi_\qq-\delta_\KK}{2}\right)=\frac{1+\cos(\phi_\qq-\delta_\KK)}{2}\\
    &\sin\left(\frac{\phi_\qq-\delta_\KK}{2}\right)\cos\left(\frac{\phi_\qq-\delta_\KK}{2}\right)=\frac{\sin(\phi_\qq-\delta_\KK)}{2}
\end{align}

In the case of  $E^{(\KK,+)*}_{\qq,\alpha}(\rr_{1})E^{(\KK,+)}_{\qq,\beta}(\rr_{2})$, we group the terms in the following way:
\begin{align}
    E^{(\KK,\pm)*}_{\qq,\alpha}(\rr_{1})E^{(\KK,\pm)}_{\qq,\beta}(\rr_{2})=&\frac{e^{i \KK\cdot \RR} e^{i \qq\cdot \RR}}{2}\left\lbrace \left[E_{\KK,1,\alpha}^{*}(\rr_1)E_{\KK,1,\beta}(\rr_1)+E_{\KK,2,\alpha}^{*}(\rr_1)E_{\KK,2,\beta}(\rr_1)\right]\right.\nonumber\\
    &-\cos(\phi_\qq-\delta_\KK)\left[E_{\KK,1,\alpha}^{*}(\rr_1)E_{\KK,1,\beta}(\rr_1)-E_{\KK,2,\alpha}^{*}(\rr_1)E_{\KK,2,\beta}(\rr_1)\right]\\
    &\left.-\sin(\phi_\qq-\delta_\KK)\left[E_{\KK,1,\alpha}^{*}(\rr_1)E_{\KK,2,\beta}(\rr_1)+E_{\KK,2,\alpha}^{*}(\rr_1)E_{\KK,1,\beta}(\rr_1)\right]\right\rbrace\nonumber
\end{align}

Putting this product in the integral $I_{\alpha,\beta}^{(\KK,+)}$, we obtain three types of integrals:
\begin{equation}
    I_{\alpha,\beta}^{(\KK,+)}=\frac{e^{i \KK\cdot\RR}}{4\omega_{A}}\left[a_{\alpha\beta}^{\KK}(\rr_1)\II_{\KK,1}^{+}-b_{\alpha\beta}^{\KK}(\rr_1)\II_{\KK,2}^{+}-c_{\alpha\beta}^{\KK}(\rr_1)\II_{\KK,3}^{+}\right]
\end{equation}
where,
\begin{align}
    a_{\alpha\beta}^\KK(\rr_1)=&E_{\KK,1,\alpha}^{*}(\rr_1)E_{\KK,1,\beta}(\rr_1)+E_{\KK,2,\alpha}^{*}(\rr_1)E_{\KK,2,\beta}(\rr_1)\\
    b_{\alpha\beta}^\KK(\rr_1)=&E_{\KK,1,\alpha}^{*}(\rr_1)E_{\KK,1,\beta}(\rr_1)-E_{\KK,2,\alpha}^{*}(\rr_1)E_{\KK,2,\beta}(\rr_1)\\
    c_{\alpha\beta}^\KK(\rr_1)=&E_{\KK,1,\alpha}^{*}(\rr_1)E_{\KK,2,\beta}(\rr_1)+E_{\KK,2,\alpha}^{*}(\rr_1)E_{\KK,1,\beta}(\rr_1)
\end{align}

and the integrals are:
\begin{align}
    \II_{\KK,1}^{+}=&\int_{0}^{q_c}\int_{0}^{2\pi}q d\phi_{\qq}dq\; \frac{e^{i q R \cos[\phi-\phi_{\qq}]}}{\omega_{A}-\omega_{D}-v q}\\
    \II_{\KK,2}^{+}=&\int_{0}^{q_c}\int_{0}^{2\pi}q d\phi_{\qq}dq\; \frac{e^{i q R \cos[\phi-\phi_{\qq}]}\cos[\phi_{\qq}-\pi/6]}{\omega_{A}-\omega_{D}-v q}\\
    \II_{\KK,3}^{+}=&\int_{0}^{q_c}\int_{0}^{2\pi}q d\phi_{\qq}dq\; \frac{e^{i q R \cos[\phi-\phi_{\qq}]}\sin[\phi_{\qq}-\pi/6]}{\omega_{A}-\omega_{D}-v q}
\end{align}

To evaluate the integral, we use the polar representation of $\RR=R(\cos(\phi),\sin(\phi))$. Additionally, we do the change of variable $\tilde{q}= vq/\delta_A$, with $\delta_A=\omega_D-\omega_A$. When $\delta_A$ is small, the upper limit for the integral in this new variable is $\tilde{q_c}=v q_c/\delta_A \gg 1$; then, we can extend this limit to infinity.
\begin{align}
    \II_{\KK,1}^{+}&=-\frac{\delta_A}{v^2}\int_{0}^{\infty}\int_{0}^{2\pi}\q d\phi_{\qq}d\q\; e^{i \q  \R \cos[\phi-\phi_{\qq}]} \frac{1}{1+\q}=-\frac{\pi\delta_{A}}{v^2}\left[\frac{2}{\R}+\pi Y_{0}(\R)-\pi H_{0}(\R)\right]\\
    \II_{\KK,2}^{+}&=-\frac{\delta_A}{v^2}\int_{0}^{\infty}\int_{0}^{2\pi}\q d\phi_{\qq}d\q\; e^{i \q  \R \cos[\phi-\phi_{\qq}]} \frac{\cos[\phi_{\qq}-\delta_\KK]}{1+\q}=\frac{i\pi^2\delta_{A}}{v^2}\cos[\delta_\KK-\phi]\left[Y_{1}(\R)+ H_{-1}(\R)\right]\\
    \II_{\KK,3}^{+}&=-\frac{\delta_A}{v^2}\int_{0}^{\infty}\int_{0}^{2\pi}\q d\phi_{\qq}d\q\; e^{i \q  \R \cos[\phi-\phi_{\qq}]} \frac{\sin[\phi_{\qq}-\delta_\KK]}{1+\q} =-\frac{i\pi^2\delta_{A}}{v^2}\sin[\delta_\KK-\phi]\left[Y_{1}(\R)+ H_{-1}(\R)\right]
\end{align}
where $\R= R\delta_A/v$, $Y_j(x)$ is the second kind Bessel function, and $H_j(x)$ is the Struve function of order $j$. For the lower band, we have the following contributions
\begin{equation}
    I_{\alpha,\beta}^{(\KK,-)}=\frac{e^{i \KK\cdot\RR}}{4\omega_{A}}\left[a_{\alpha\beta}^{\KK}(\rr_1)\II_{\KK,1}^{-}+b_{\alpha\beta}^{\KK}(\rr_1)\II_{\KK,2}^{-}+c_{\alpha\beta}^{\KK}(\rr_1)\II_{\KK,3}^{-}\right]
\end{equation}
where the integrals are:
\begin{align}
    \II_{\KK,1}^{-}&=-\frac{\delta_A}{v^2}\int_{0}^{\infty}\int_{0}^{2\pi}\q d\phi_{\qq}d\q\; e^{i \q  \R \cos[\phi-\phi_{\qq}]} \frac{1}{1-\q}\nonumber\\
    &=-\frac{\pi\delta_{A}}{v^2}\left[\frac{2}{\R}-\pi Y_{0}(\R)-\pi H_{0}(\R)+2i \pi J_0(\R)\right]\, ,\\
    \II_{\KK,2}^{-}&=-\frac{\delta_A}{v^2}\int_{0}^{\infty}\int_{0}^{2\pi}\q d\phi_{\qq}d\q\; e^{i \q  \R \cos[\phi-\phi_{\qq}]} \frac{\cos[\phi_{\qq}-\delta_\KK]}{1-\q}\nonumber\\
    &=-\frac{i\pi^2\delta_{A}}{v^2}\cos[\delta_\KK-\phi]\left[Y_{1}(\R)- H_{-1}(\R)-2iJ_1(\R)\right]\, ,\\
    \II_{\KK,3}^{-}&=-\frac{\delta_A}{v^2}\int_{0}^{\infty}\int_{0}^{2\pi}\q d\phi_{\qq}d\q\; e^{i \q  \R \cos[\phi-\phi_{\qq}]} \frac{\sin[\phi_{\qq}-\delta_\KK]}{1-\q}\nonumber\\
    &=\frac{i\pi^2\delta_{A}}{v^2}\sin[\delta_\KK-\phi]\left[Y_{1}(\R)-H_{-1}(\R)-2iJ_1(\R)\right]\, ,
\end{align}
$J_j(x)$ is the first kind Bessel function of order $j$. These integrals have poles, so we add a small imaginary part in the denominator of the integrand and make the integration using the  Sokhotski–Plemelj theorem for the real line:
$$\lim_{\zeta\rightarrow 0^+}\int_{a}^{b}\frac{f(x)}{x\pm i\zeta}dx= \mathcal{P}\left[\int_{a}^{b}\frac{f(x)}{x}\right]\mp i\pi f(0)$$
where $\mathcal{P}$ represents the Cauchy principal value, and $a<0<b$. The sign $+(-)$ in the denominator finally results in the first (second) kind Hankel function, representing the outgoing (incoming) waves. In contrast with Ref.~\citenum{Perczel2020a}, we have chosen the first kind Hankel function, so that the imaginary part at the $R=0$ position is positive.

Similar to $\KK$, we can write the contribution for $\KK'$ in terms of three integrals:
\begin{eqnarray}
I_{\alpha,\beta}^{(\KK',+)}&=&\frac{e^{i \KK'\cdot\RR}}{4\omega_{A}}\left[a_{\alpha\beta}^{\KK'}(\rr_1)\II_{\KK',1}^{+}-b_{\alpha\beta}^{\KK'}(\rr_1)\II_{\KK',2}^{+}+c_{\alpha\beta}^{\KK'}(\rr_1)\II_{\KK',3}^{+}\right]\\
I_{\alpha,\beta}^{(\KK',-)}&=&\frac{e^{i \KK'\cdot\RR}}{4\omega_{A}}\left[a_{\alpha\beta}^{\KK'}(\rr_1)\II_{\KK',1}^{-}+b_{\alpha\beta}^{\KK'}(\rr_1)\II_{\KK',2}^{-}-c_{\alpha\beta}^{\KK'}(\rr_1)\II_{\KK',3}^{-}\right]
\end{eqnarray}
where the explicit forms of $a_{\alpha\beta}^{\KK'}$,
$b_{\alpha\beta}^{\KK'}$, $C_{\alpha\beta}^{\KK'}$ and $I_{\KK',m}^\pm$ are analogous to the case of $\KK$, the only difference is to change $\KK$ to $\KK'$.

Adding all the contributions we obtain
\begin{align}
G_{\alpha\beta}(\rr_1;\RR)\approx \frac{\sqrt{3}a^2c^2\delta_A}{16\, \omega_A v^2}&\left\lbrace i H_0^{(1)}(R\delta_A/v) \left[\mathcal{A}^{\KK}_{\alpha\beta}(\rr_1) e^{i\KK\cdot\RR}+\mathcal{A}^{\KK'}_{\alpha\beta}(\rr_1) e^{i\KK'\cdot\RR}\right] \right.\nonumber\\
&\left.-H_1^{(1)}(R\delta_A/v)\mathcal{B}^{\KK}_{\alpha\beta}(\rr_1,\phi)e^{i\KK\cdot\RR}-H_1^{(1)}(R\delta_A/v)\mathcal{B}^{\KK'}_{\alpha\beta}(\rr_1,\phi)e^{i\KK'\cdot\RR}·\right\rbrace
~\label{SI-Ec: Gab}
\end{align}
where $H_j^{(1)}(x)$ is the first kind Hankel function of order $j$, and the information of the electric field modes are given by the $\mathcal{A}$ and $\mathcal{B}$ coefficients which read:
\begin{align}
    &\mathcal{A}_{\alpha\beta}^{\KK^{(')}}(\rr_1)=a_{\alpha\beta}^{\KK^{(')}}(\rr_1)\\
    &\mathcal{B}_{\alpha\beta}^{\KK}(\rr_1,\phi)=b_{\alpha\beta}^{\KK}(\rr_1)\cos[\delta_{\KK}-\phi]-c_{\alpha\beta}^{\KK}(\rr_1)\sin[\delta_{\KK}-\phi]\\
    &\mathcal{B}_{\alpha\beta}^{\KK'}(\rr_1,\phi)=b_{\alpha\beta}^{\KK'}(\rr_1)\cos[\delta_{\KK'}-\phi]+c_{\alpha\beta}^{\KK'}(\rr_1)\sin[\delta_{\KK'}-\phi]
\end{align}
remembering $\delta_{\KK^{(')}}=\pm \pi/6$. Additionally, $\KK$ and $\KK'$ are conected by a rotation and inversion, so the fields have the following symmetry relations:
\begin{equation}
    \begin{pmatrix}
        E_{\KK',1,\alpha(\beta)}(\rr_1)\\
        E_{\KK',2,\alpha(\beta)}(\rr_1)
    \end{pmatrix}=\begin{pmatrix}
        1&& 0\\
        0&& -1
    \end{pmatrix}\begin{pmatrix}
        \cos(\pi/3)&& -\sin(\pi/3)\\
        \sin(\pi/3)&& \cos(\pi/3)
    \end{pmatrix}\begin{pmatrix}
        E_{\KK,1,\alpha(\beta)}^{*}(\rr_1)\\
        E_{\KK,2,\alpha(\beta)}^{*}(\rr_1)
\end{pmatrix}
\end{equation}
using these symmetry relations, it is possible to link the $a_{\alpha\beta}^{\KK'}(\rr_1)$, $b_{\alpha\beta}^{\KK'}(\rr_1)$ and $c_{\alpha\beta}^{\KK'}(\rr_1)$, with their analogous in the symmetry point $\KK$, as follow:
\begin{align}
      a_{\alpha\beta}^{\KK'}(\rr_1)=&a_{\alpha\beta}^{\KK *}(\rr_1)\label{SI-ec:aKp}\\
      b_{\alpha\beta}^{\KK'}(\rr_1)=&-\frac{1}{2}b_{\alpha\beta}^{\KK *}(\rr_1)+\frac{\sqrt{3}}{2}c_{\alpha\beta}^{\KK *}(\rr_1)\label{SI-ec:bKp}\\
      c_{\alpha\beta}^{\KK'}(\rr_1)=&\frac{\sqrt{3}}{2}b_{\alpha\beta}^{\KK *}(\rr_1)+\frac{1}{2}c_{\alpha\beta}^{\KK *}(\rr_1)\label{SI-ec:cKp}
\end{align}

In the numerical calculation, these relations are not exactly satisfied, especially close the discontinuities in the dielectric index due where the convergence of GME method is more challenging. This induces a small numerical error when calculating the Green function, that leads to unphysical results, like having a non-zero value of the imaginary part in the limit $R\rightarrow 0$ (single point Green function) when it should be exactly $0$ due to the vanishing density of states.  To circumvent this problem, we do not calculate the $a_{\alpha\beta}^{\KK'}(\rr_1)$, $b_{\alpha\beta}^{\KK'}(\rr_1)$ and $c_{\alpha\beta}^{\KK'}(\rr_1)$ directly from the GME method, but instead we use the results of $a_{\alpha\beta}^{\KK}(\rr_1)$, $b_{\alpha\beta}^{\KK}(\rr_1)$ and $c_{\alpha\beta}^{\KK}(\rr_1)$ to calculate their analogous at $\KK'$ using the Eqs.~\ref{SI-ec:aKp}-\ref{SI-ec:cKp}.

\subsection{Dirac-Photon-mediated interactions: additional results}

As shown in Eqs.~\eqref{SI-Ec: Gab} the dependence of the Green function on the distance between emitters stems from the combination of Hankel functions $H_j^{(1)}(x)$. In the case where the emitters are tuned close to the Dirac point, i.e., $\delta_A\rightarrow 0$, these functions display an asymptotic scaling with:
\begin{align}
H_0^{(1)}(x)&\approx 1+i\frac{2}{\pi}\left[\log(x/2)+ \gamma_E/\pi\right]\,,\\
H_{1}^{(1)}(x)&\approx \frac{x}{2}-i\frac{2}{\pi x}\,,
\end{align}
for $|x|\ll 1$, where $\gamma_E$ is the Euler constant. Depending on the emitters' position, the coefficients $\mathcal{A}$ and $\mathcal{B}$ accompanying will have different values, and thus lead to a different combination. As shown in the main text for  $G_{xx}(\rr_1;\RR)$ and $G_{\sigma_+\sigma_+}(\rr_1;\RR)$, the resulting spatial decay can be always shown to be oscillating and with a power-law decay. Thus, the Green function can be fitted to a power-law envelope function $\propto 1/|\RR|^\gamma$, where the $\gamma$ gives us a quantifier of the effective range of the interactions emerging from the combination of these Hankel functions. Additionally to the range of interactions, another relevant magnitude is their strength; we quantified this by the absolute value of the two-point Green function between two nearest-neighbouring emitters, $|G_{\alpha\beta}(\rr_1)^{(1)}|$. Finally, it is essential to characterize their coherent (incoherent) nature; to do this, we use the quantifier $W_1(\rr_1)$ defined in the main text.

Here, we display some additional results for other components that are not shown in the main text. First, we present the results for linear polarization, and second for circular polarization.

\subsubsection{Linear polarization}

\begin{figure}[tb]
    \centering
    \includegraphics[scale=0.46]{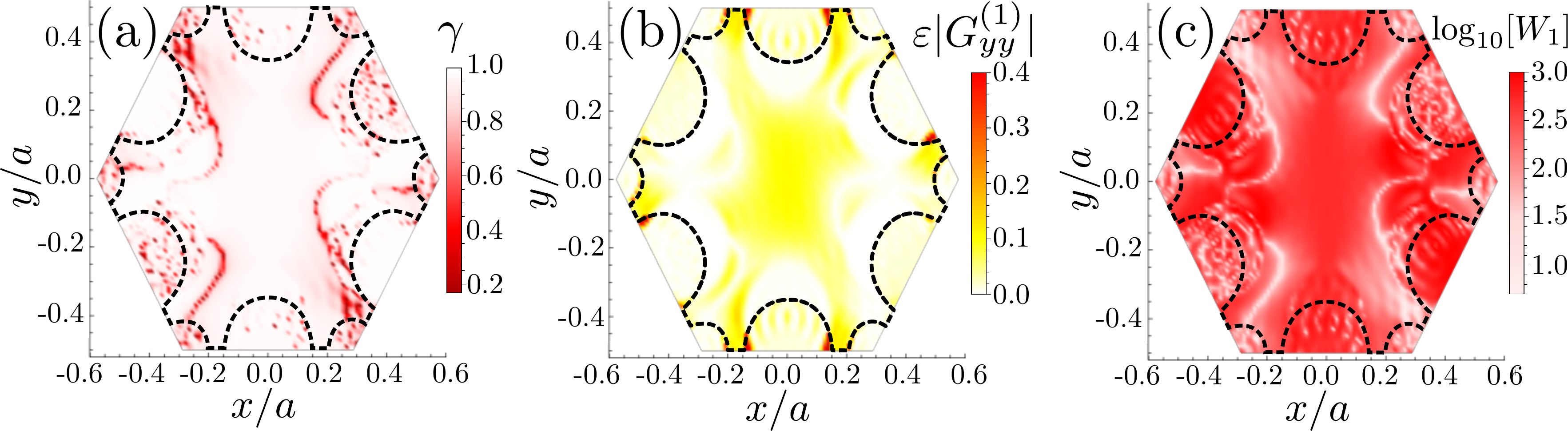}
    \caption{Behaviour of $G_{yy}$ for the direction of $\phi=\pi/6$. The parameters of the structure slab and emitters are the same as in Fig.~\ref{SI-fig:Lukin_comparison}, the transition wavelength of the emitter is $738$ nm and $\delta_A/2\pi=19.5$ GHz. (a) Behaviour of the decay ($\gamma$) at all positions of the unit cell, in (b) we show the strength of the interaction with the first neighbour multiplied by $\varepsilon$ and (d) show the coherent(incoherent) nature of the interaction by plotting $W_1(\rr_1)$ (see main text) in logarithmic scale, defined here for $G_{yy}^{(1)}(\rr_1)$.}
    \label{SI-fig:Gyy}
\end{figure}

\begin{figure}[tb]
    \centering
    \includegraphics[scale=0.46]{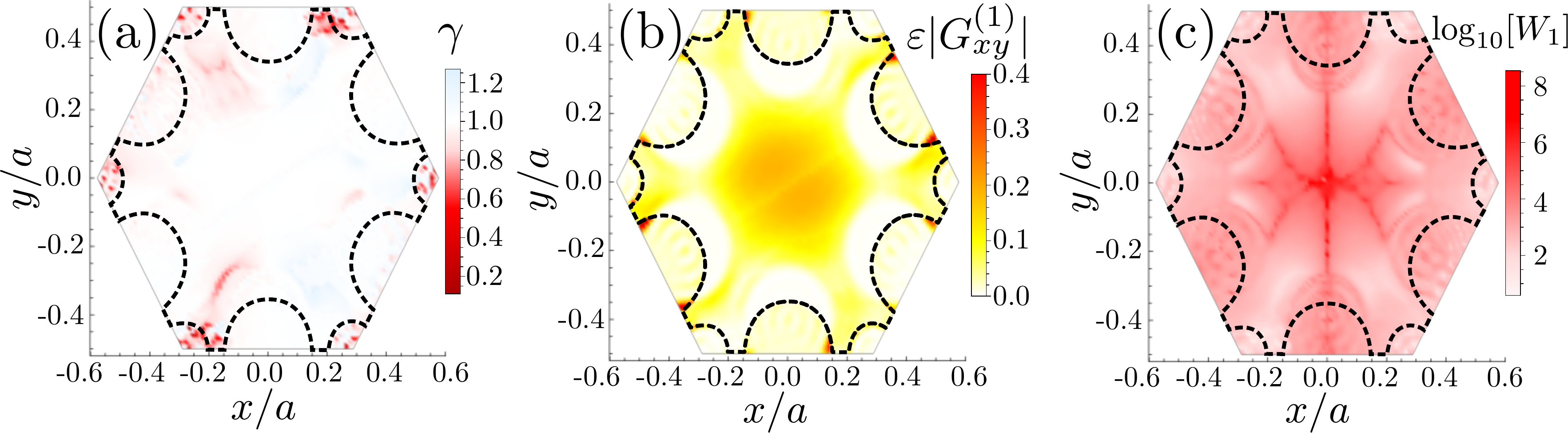}
  \caption{Behaviour of $G_{xy}$ for the direction of $\phi=\pi/6$, the parameters of the structure slab and emitters are the same as in Fig. \ref{fig:Gxx} of the main text. (a) Behaviour of the decay ($\gamma$) at all position in unit cell, in (b) we show the strength of the interactions with the first neighbour multiplied by $\varepsilon$ and (d) show the coherent(incoherent) nature of the interactions by $W_1(\rr_1)$ (see main text) in logarithmic scale, define here for $G_{xy}^{(1)}(\rr_1)$.}
    \label{SI-fig:Gxy}
\end{figure}
For linear polarization, we show the interaction parameters for the components $G_{yy}$ in Fig. \ref{SI-fig:Gyy} and $G_{xy}$ in Fig. \ref{SI-fig:Gxy}. Figs \ref{SI-fig:Gyy}(a) and \ref{SI-fig:Gxy}(a) present the $\gamma$ parameter for each component. Similar to $G_{xx}$ in the main text, we observe that $\gamma$ is close to $1$ in almost all the unit cell for $G_{yy}$ and $G_{xy}$, except for some regions in red that have $\gamma<1$ with longer-ranged interactions. In the case of the component $G_{xy}$, some small regions where the $\gamma$ is slightly greater than one are in light-blue color in Fig.~\ref{SI-fig:Gxy}(a). Remembering the trade-off between the range and strength of the interactions discussed in the main text, we also characterize the strength in Figs~\ref{SI-fig:Gyy}(b) and~\ref{SI-fig:Gxy}(b) for both components. We find similar results as in the main text: the regions with longer-range interactions have less interaction strength. However, it is possible to choose the position that tunes an interaction with $\gamma<1$ and relevant strength. We also observe that positions with maximum strength are between the air holes for emitters inside the dielectric and close to the edge of holes for emitters lying in the air. For cross-polarization ($G_{xy}$ component), the region around the center has a good strength similar to the maximum at the optimal position. 
Finally, in Figs~\ref{SI-fig:Gyy}(c) and \ref{SI-fig:Gxy}(c), we characterize the nature of the interactions by plotting in logarithmic scale $W_{1}$, finding that the interactions are coherent along most of the unit cell, as also occurred for the $G_{xx}$ component shown in the main text.

\subsubsection{Circular Polarization}

\begin{figure}[tb]
    \centering
    \includegraphics[scale=0.46]{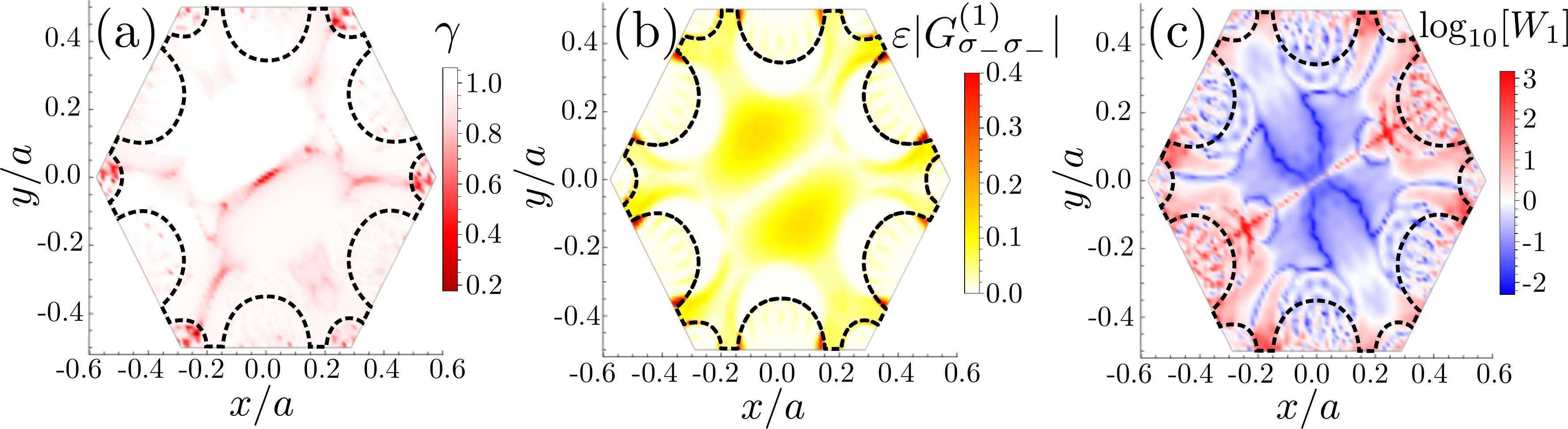}
  \caption{Behaviour of $G_{\sigma_-\sigma_-}$ for the direction of $\phi=\pi/6$. The parameters of the structure slab and emitters are the same as in Fig.~\ref{fig:Gxx} of the main text. (a) Behaviour of the decay ($\gamma$) at all position in unit cell, in (b) we show the strength of the interaction with the first neighbour multiplied by $\varepsilon$ and (d) show the coherent(incoherent) nature of the interaction by $W_1(\rr_1)$ (see main text) in logarithmic scale, defined here for $G_{\sigma_-\sigma_-}^{(1)}(\rr_1)$.}
    \label{SI-fig:Gmm}
\end{figure}

\begin{figure}[tb]
    \centering
    \includegraphics[scale=0.46]{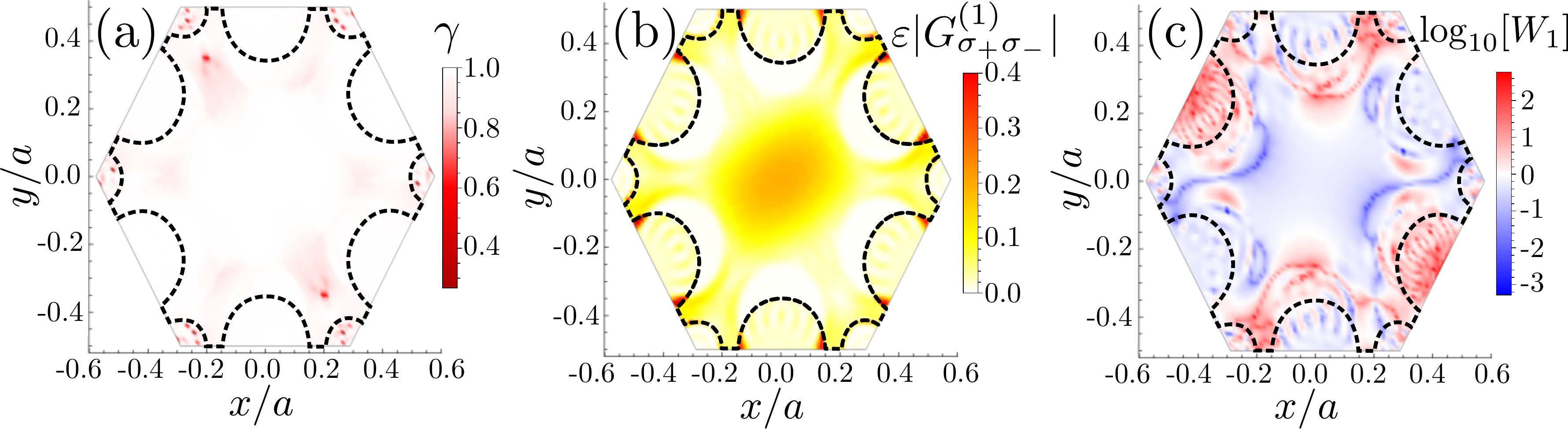}
  \caption{Behaviour of $G_{\sigma_+\sigma_-}$ for the direction of $\phi=\pi/6$, the parameters of the structure slab and emitters are the same as in Fig.~\ref{fig:Gxx} of the main text. (a) Behaviour of the decay ($\gamma$) at all position in unit cell, in (b) we show the strength of the interaction with the first neighbour multiplied by $\varepsilon$ and (d) show the coherent(incoherent) nature of the interaction by $W_1(\rr_1)$ (see main text) in logarithmic scale, define here for $G_{\sigma_+\sigma_-}^{(1)}(\rr_1)$.}    \label{SI-fig:Gpm}
\end{figure}

For the case of circularly polarized transitions, we also find a trade-off between the range and strength for $G_{\sigma_-\sigma_-}$ and $G_{\sigma_+\sigma_-}$ components. This is illustrated in Figs.~\ref{SI-fig:Gmm}(a-b) and \ref{SI-fig:Gpm}(a-b), for each polarization, respectively. Although we find the optimal position very close to the holes for emitters inside the dielectric, some regions close to the center have large coupling strength for both components. As it occurs for the $G_{\sigma_+\sigma_+}$ component shown in the main text, the center is not an optimal position to couple emitters with $\sigma_-$ polarization, since the strength at this position is approximately zero. This does not occur for the cross polarization situation, i.e., $G_{\sigma_+\sigma_-}$ component. In the case of emitters in air holes, the maximum strength is found very close to the interface air-dielectric for all the components in both polarization bases. One important difference with respect to the linearly polarized cases is that the nature of the interactions can change from being coherent to incoherent, as shown already for the main text for $G_{\sigma_+\sigma_+}$. This also occurs for the other circularly polarized components as shown in detail in Figs.~\ref{SI-fig:Gmm}(c) and~\ref{SI-fig:Gpm}(c).


\end{widetext}

\bibliographystyle{apsrev4-1}
\bibliography{references}

\end{document}